\documentclass[prd,aps,floats,twocolumn,nofootinbib]{revtex4-1}
\usepackage{slashed}
\usepackage{mathtools}
\usepackage{amsfonts}
\usepackage{amssymb}
\usepackage{epsfig}

\begin{document}

\newcommand{\m}[1]{\mathcal{#1}}
\newcommand{\nn}{\nonumber}
\newcommand{\ph}{\phantom}
\newcommand{\eps}{\epsilon}
\newcommand{\be}{\begin{equation}}
\newcommand{\ee}{\end{equation}}
\newcommand{\bea}{\begin{eqnarray}}
\newcommand{\eea}{\end{eqnarray}}
\newtheorem{conj}{Conjecture}

\newcommand{\plk}{\mathfrak{h}}
\newcommand{\bb}{\bar b}


\title{Unitarity and Vilenkin's wave functions}
\date{ }

\author{Bruno Alexandre}
\author{Raymond Isichei}
\author{Jo\~{a}o Magueijo}
\email{j.magueijo@imperial.ac.uk}

\affiliation{Theoretical Physics Group, The Blackett Laboratory, Imperial College, Prince Consort Rd., London, SW7 2BZ, United Kingdom}

\begin{abstract}
It is remarkably difficult to reconcile unitarity and Vilenkin's wave function.
For example, the natural conserved inner product found in quantum unimodular gravity 
applies to the Hartle-Hawking wave function, but fails for its Vilenkin counterpart. We diagnose this failure from different angles (Laplace transform instead of Fourier transform, non-Hermiticity of the Hamiltonian, etc) to conclude that ultimately it stems from allowing the connection to become imaginary in a section of its contour. In turn this is the unavoidable consequence of representing the Euclidean theory as an imaginary image within a fundamentally Lorentzian theory.  It is nonetheless possible to change the underlying theory and replace the 
connection's foray into the imaginary axis by an actual signature change (with the connection, action and Hamiltonian remaining real). The structural obstacles to unitarity are then removed, but special care must still be taken, because the Euclidean theory {\it a priori} has boundaries, so that appropriate boundary conditions are required for unitarity. Reflecting boundary conditions would reinstate a Hartle-Hawking-like solution in the Lorentzian regime. To exclude an incoming wave in the Lorentzian domain one must allow a semi-infinite tower of spheres in the Euclidean region, wave packets travelling through successive spheres for half an eternity in unimodular time. Such ``Sisyphus'' boundary condition no longer even vaguely resembles Vilenkin's original proposal. 
\end{abstract}

\maketitle

\section{Introduction}

The fate of unitarity in the deep quantum gravity/cosmology regime is a matter of debate. Preserving unitarity may be seen as a challenge to be faced by any successful quantum gravity theory~\cite{LeeReview}; it may also be a mirage: an approximate concept valid only in the semi-classical limit, to be abandoned as soon as we plunge into the Planck epoch~\cite{Vil-interpretation}. Without wanting to take sides on this dispute, in this paper we investigate the practicalities of implementing unitarity in the context of Vilenkin's proposal for the wave function of the Universe~\cite{vil-PRD,Vil-review}.  

We base our efforts on recent work~\cite{HHpackets} showing how unimodular gravity~\cite{unimod1,unimod,UnimodLee1,alan,daughton,sorkin1,sorkin2} may be used to improve the physical interpretation of the Hartle-Hawking proposal~\cite{HH} and its compliance with unitarity. 
In the formulation of~\cite{unimod},  the unimodular gravity extension demotes the cosmological constant $\Lambda$ from a pre-given parameter to a classical constant of motion, with a conjugate momentum providing a physical measure of time: unimodular time ~\cite{unimod,Bombelli,UnimodLee2}. Quantum mechanically, this permits superpositions of Hartle-Hawking wave functions with different $\Lambda$ to form wave packets~\cite{UnimodLee2,HH}. This suggests an inner product, with respect to which these packets and other physical states are normalizable. More importantly, the evolution in unimodular time is unitary with respect to this inner product. In view of this conservation, it is possible to set up a probability interpretation valid beyond the semi-classical limit.

None of this applies to Vilenkin's ``tunneling'' wave function~\cite{vil-PRD,Vil-review},  which seems at odds with any implementation of unitarity and probability, a feature which appears to be intentional~\cite{Vil-interpretation}.  
In this paper we re-examine the tunneling wave function, first from the point of view of the connection, rather than the metric representation (Section~\ref{first}), then from the additional perspective of the unimodular extension (Sections~\ref{HH} and \ref{VLap}). A number of promising first results are found in Section~\ref{first} by translating the monochromatic (fixed $\Lambda$) solutions from metric to connection representation. However, any attempts to lift the procedure in~\cite{HHpackets} (reviewed in Section~\ref{HH}) to the Vilenkin setup fail. We are thus unable to provide an equivalent definition of probability. We diagnose this failure from different angles. In Section~\ref{VLap} we note that Vilenkin's proposal requires the connection to stray off the real line~\cite{CSHHV}, leading to the unavoidable use of the Laplace transform instead of the Fourier transform. The inverse Laplace transform then places the probability on the Hartle-Hawking contour only. In Section~\ref{nail} we blame something more fundamental: the non-Hermiticity of the Hamiltonian. 

\begin{figure}
\centering
\begin{tabular}{c c}
    \includegraphics[scale=0.064]{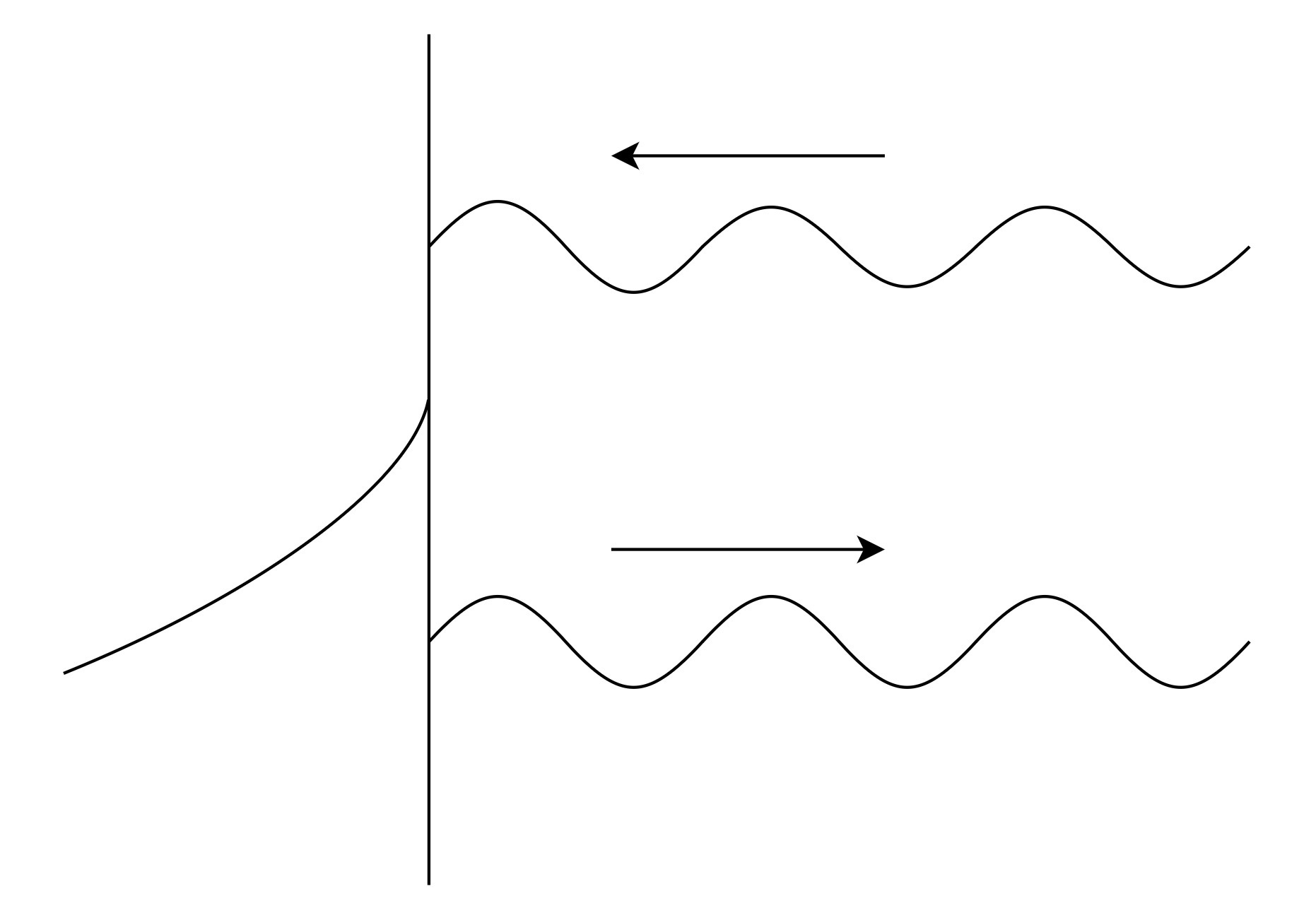} & \includegraphics[scale=0.064]{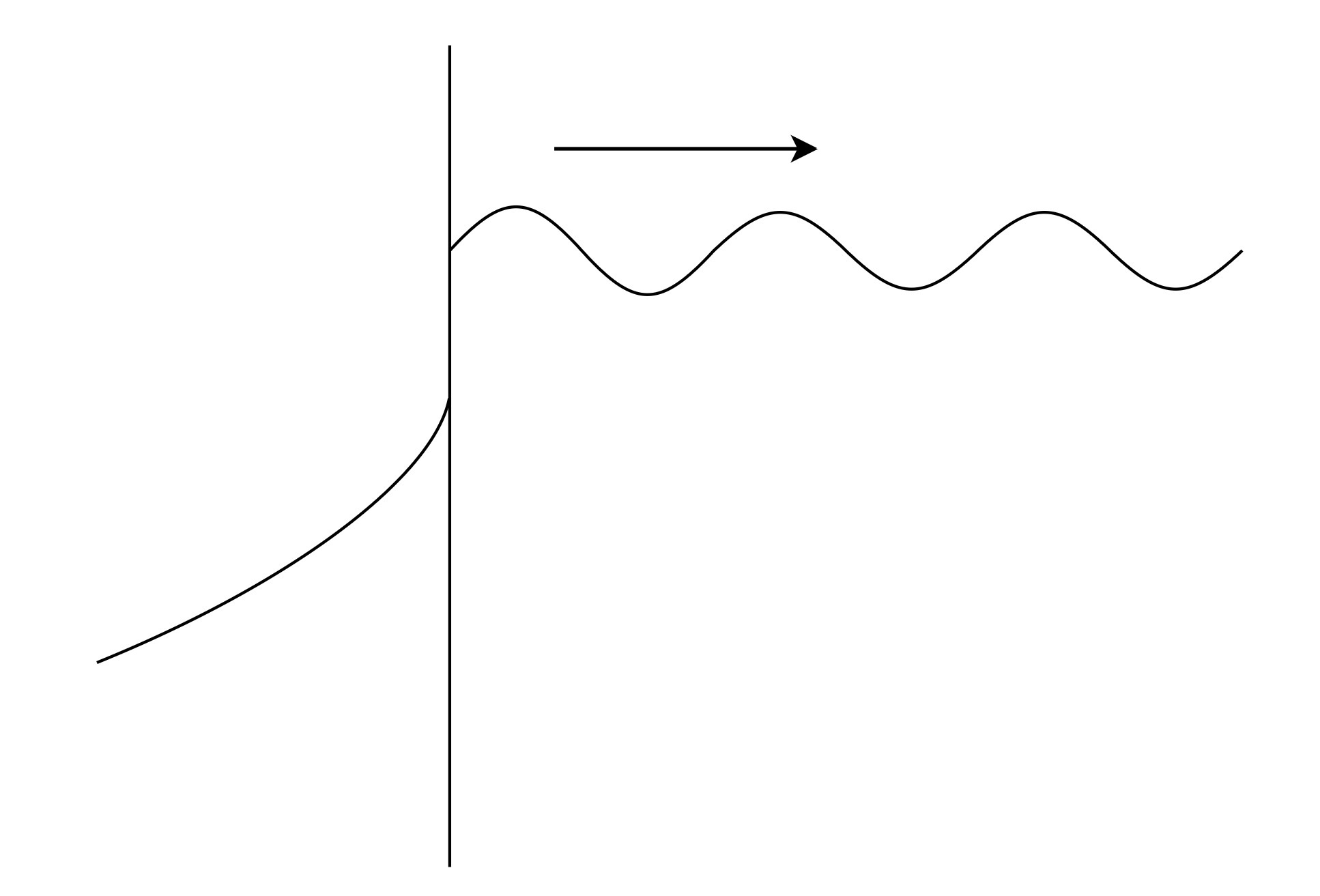}  
\end{tabular}
\caption{In metric space, the Hartle-Hawking wave function (left) can be seen as the reflection of an incident wave off a wall, resulting in a reflected wave and an evanescent wave penetrating the wall. The Vilenkin wave function (right) amounts to an evanescent wave in the classically forbidden region spitting out an outgoing wave.} 
\label{cartoons}
\end{figure}

This failure is physically intuitive.  The Hartle-Hawking and Vilenkin proposals loosely correspond to the two situations depicted in Fig.~\ref{cartoons}. In metric space the Hartle-Hawking wave function (left) can be seen as the reflection of an incident wave off a wall, resulting in a reflected wave and an evanescent wave penetrating the classically forbidden region. The Vilenkin wave function (right) amounts to an evanescent wave spitting out an outgoing wave, with the conspicuous absence of an incident wave.  This is hardly an acceptable physical situation under our standard intuition on conservation of probability. But quantum gravity is not {\it standard}, and two new ingredients arise from the use of the connection representation instead of the metric representation,  and  the use of unimodular time to resolve a problem (the problem of time) not usually found in standard quantum mechanics. In Sections~\ref{VLap} and~\ref{nail}, however, we show that these additions do not qualitatively change anything. 

Mathematically, however, there is hope. Both our diagnoses stem from allowing the connection to become imaginary in part of its domain~\cite{CSHHV}.  This is the unavoidable consequence of representing the Euclidean theory as an imaginary {\it image} within a fundamentally Lorentzian theory (Section~\ref{EucImage}). But 
what if the Universe undergoes a signature change as part of the action, so that the classical trajectories themselves follow a path from the Euclidean sphere to the Lorentzian de Sitter space time (Section~\ref{EucSignChange})? This is totally different from the instanton obtained by allowing the connection to go over the imaginary domain. With this assumption, the connection, action and Hamiltonian remain real, so that the structural obstacles to unitarity are removed. 

However, as described in Section~\ref{EucQuantum}, we still have work to do because the theory has boundaries (a 4-sphere is finite). In order to obtain an outgoing without incoming wave we must discount a reflection at the South pole and indeed continue the Euclidean half sphere with an infinite tower of spheres, playing the role of the proverbial turtles in ancient cosmologies (Sections~\ref{Sisyphus} and \ref{other}). This continuation does not rely on a theory of gravity allowing degenerate metrics, and so a {\it classical} gluing of the North pole of one sphere with the South pole of the one above. The wave function never peaks at the gluing point (as our central Fig~\ref{figprob} illustrates). As unimodular time progresses a superposition of peaks at non-singular points in adjoining spheres is found, with a gradual handover between the two. 

This phenomenon is reminiscent of the singularity resolution described in~\cite{GielenSing}, and it may help stave off runaway perturbation instabilities at North/South joining points, as described in Refs.~\cite{JL1,JL2,JL3,Witten,JeanLuc}, specifically~\cite{JeanLuc} (but see also~\cite{Vilreply1,Vilreply2}). Indeed this possibility can be seen as a major motivation for our work. If the fixed-$\Lambda$ monochromatic background is deemed unphysical from the unimodular perspective, one might not be too surprised that the fluctuations around it are unbounded. The construction of physical packets could open up the doors to a solution to these instabilities from the unimodular perspective. The fact that we have to modify the original Vilenkin proposal so much before this can be implemented shows how problematic the proposal is. It remains to be seen whether the fact that the Universe is never
semiclassical at the problematic points is sufficient to resolve the problem.

\section{First hopes}\label{first}
The first ingredient in this paper is the use of the connection representation, rather than the more frequently used metric representation. Except for very standard situations, the choice of representation in the quantum theory is not innocuous. It can lead to inequivalent theories: for example with different natural inner products and probability interpretations. It may also shed new light on boundary conditions. The reason why 
the position representation is usually favoured in standard Quantum Mechanics is that it is physically clearer for defining boundary conditions.  In Quantum
Gravity it is far from obvious which representation should receive primacy in this respect. The first hope in this paper is to  reassess the chasm between the Hartle-Hawking and Vilenkin's choices from the point of view of the connection representation.


At first sight, the dictionary between metric and connection representations is straightforward~\cite{CSHHV}. Recall that the action can be written as:
\bea\label{ECaction}
S_0&=&\frac{3 V_c}{8\pi G} \int dt \bigg(  a^{2} \dot{b}
+ Na\bigg(b^2+k-\frac{\Lambda}{3}a^{2}\bigg) \bigg).
\eea
where $a$ is the expansion factor, $b$ is the only minisuperspace connection variable (an off-shell version of the Hubble parameter, since $b= \dot a$ on-shell, if there is no torsion), $k$ is the spatial curvature, $N$ is the lapse function and $V_c=\int d^3 x$ is the comoving volume of the region under study, assumed finite throughout this paper (in the quantum cosmology classical literature one usually chooses $k=1$ and $V_c=2\pi^2$; see~\cite{Afshordi} for a discussion of the criteria for the choice of $V_c$). 
Throughout this paper we will assume $k>0$ and $\Lambda>0$ (we do not set $k=1$ since the comoving volume $V_c$ is left as an open scale). 

Upon quantization this action implies 
the complementarity relation:
\bea\label{commutator}
\left[\hat b,\hat{ a^2}\right]&=&\frac{i l_P^2}{3 V_c}=i\plk 
\eea
so that the wave functions in the 2 representations are just Fourier duals:
\bea\label{FT}
\psi_{a^2}(a^2)&=&
\int \frac{db }{\sqrt{2\pi\plk }} e^{-\frac{i}{\plk }a^2 b}\psi_b(b).
\eea
This is vindicated by the concrete solutions.
The connection space solution is the Chern-Simons-Kodama state:
\be\label{CS}
\psi_s(b,\Lambda)=\psi_{CS}={\cal N}\exp{\left[i \frac{9V_c}{\Lambda l_P^2}
\left(\frac{b^3}{3}+bk\right)
\right]}.
\ee
Inserting in (\ref{FT}) we find the integral representation of the Airy functions making up the Hartle-Hawking and Vilenkin wave functions, with the proviso that the choice of contour in $b$ space determines whether Hartle-Hawking (real line) or V (positive real line and negative imaginary line) is obtained (see~\cite{CSHHV} for details).

But is there more to it? What if we reassessed 
the issue of boundary conditions and probability interpretation starting from the connection representation? In the metric representation there is a physical picture of the Universe  tunnelling ``out of nothing''. In this language the ``nothing'' is the point $a=0$, so what is its counterpart in the connection representation? The Hamiltonian constraint:
\be\label{Hamconst}
H=-(b^2+k)+\frac{\Lambda}{3} a^2=0
\ee
implies that {\it classically} $a=0$ is equivalent to $b=\pm i\sqrt k $ (with $k>0$, as will be assumed throughout this paper). 
These points are absent from the contour leading to the Hartle-Hawking wave function (the real line), but one of them:
\be\label{bnought}
b=b_\emptyset= - i\sqrt k 
\ee  
lies on the contour leading to the V wave function (specifically the negative imaginary axis). What is so special about this point for the Chern-Simons wave function? 

\begin{figure}
\centering
\includegraphics[scale=0.55]{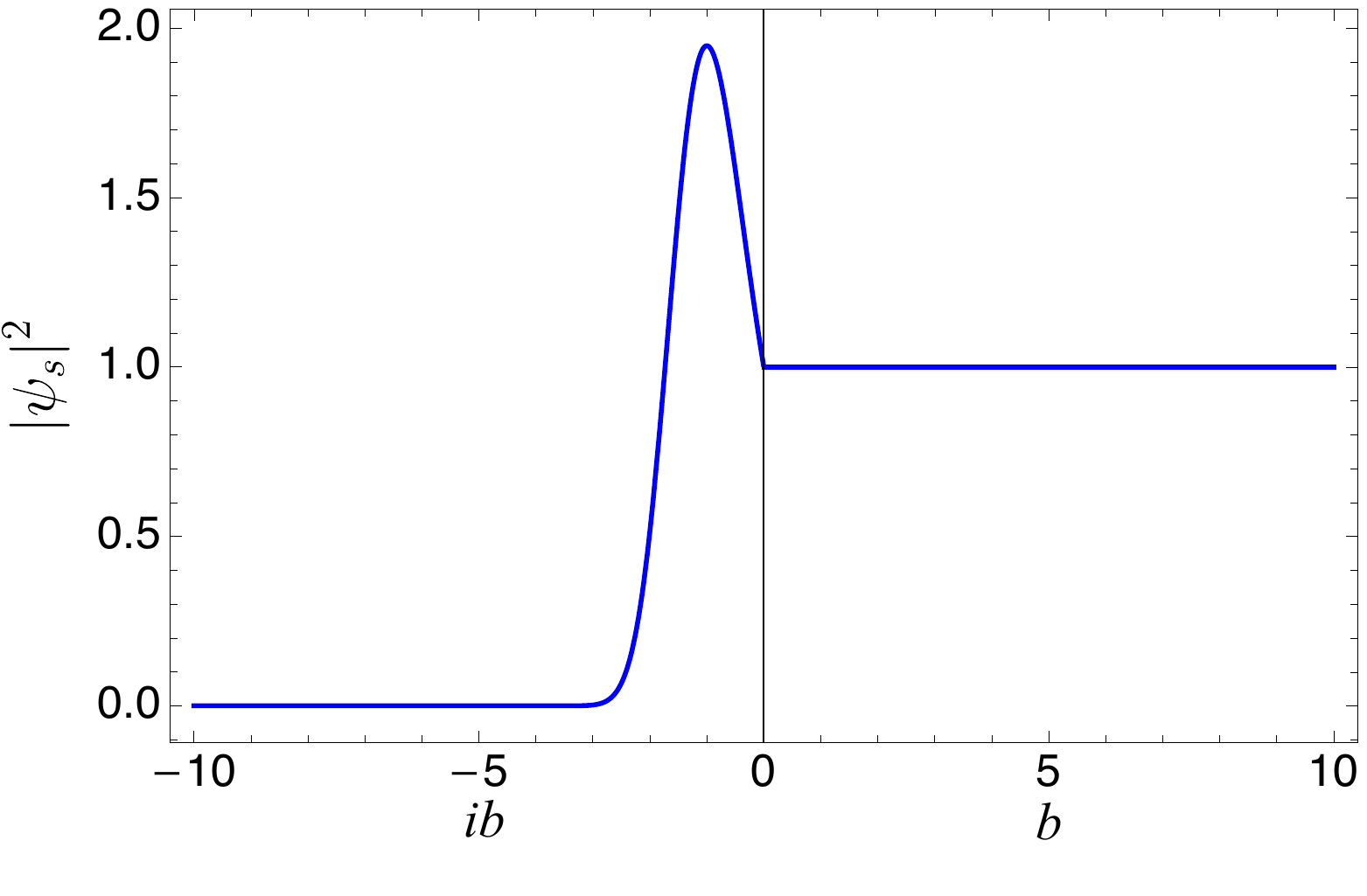}
\caption{Typical shape of $|\psi_s|^2$ in connection space along the V contour. Whereas the positive axis corresponds to the positive real line, the negative axis is on the negative {\it imaginary} line, as per the prescription detailed in~\cite{CSHHV}. There is a maximum at $b_\emptyset= - i\sqrt k$. The ratio between  the height of the (real $b$) plateau and this maximum coincides with the probability of nucleation as evaluated in~\cite{vil-PRD,Vil-review}.} 
\label{fig1}
\end{figure}

At this point it is easy to be beguiled by an interesting coincidence. In the metric representation, the probability of tunneling out of $a=0$ is is obtained via a Klein-Gordon current interpretation of probability~\cite{vil-PRD,Vil-review}, and when applied to the V wave function gives:
\begin{equation}\label{probV}
    j^a \approx \exp{\left[-\frac{12 V_c k^\frac{3}{2}}{l_P^2 \Lambda} \right]}.
\end{equation}
(we have adapted the standard result to our conventions, as in~\cite{Albertini}). 
But we could also take the Chern-Simons-Kodama state, plot $|\psi|^2$
along the contour leading to the V wave function 
to find that this is flat in along the the positive real $b$, but not in the imaginary section of the contour. Indeed $|\psi|^2$ has a maximum at $b=b_\emptyset$, tailing off along the negative imaginary axis towards $-\infty$ on one side, and connecting with the plateau along the positive real line (the classical region) on the other side 
(see Fig.\ref{fig1}). Computing the ratio between the classical plateau and this maximum we get:
\begin{equation}
    \frac{|\psi|^2_{class}}{|\psi|^2_{\emptyset}}\approx \exp{\left[-\frac{12 V_c k^\frac{3}{2}}{l_P^2 \Lambda} \right]}.
\end{equation}
i.e. precisely (\ref{probV}) (numerical factor of 12 included). This is a remarkable coincidence, strongly pointing to a reinterpretation of the standard metric/KG current result from the alternative connection point of view.


Unfortunately, none of this is left standing upon closer mathematical scrutiny. First of all, what right to we have to compute probabilities with Born's $|\psi|^2$ prescription? How does this tally with unitarity?\footnote{The ``coincidence'' highlighted in this Section is related to the fact that even in the metric representation the Klein-Gordon result can also be obtained by a basic ratio of Born factors: see Eq.(3.17) in~\cite{Vil-review}.}


\section{Unitarity and the Unimodular extension}\label{HH}

At this point it is useful to review recent progress on the issue of time, normalizability and unitarity in the context of the Hartle-Hawking wave function. As shown in~\cite{HHpackets}, by extending the theory to unimodular gravity, we can obtain both a physical time variable and a natural inner product, to enforce unitarity.  

\subsection{The unimodular extension}
We use the Henneaux and Teitelboim formulation of ``unimodular'' gravity~\cite{unimod}, where one adds to a ``base action'' $S_0$ (e.g. standard General Relativity)  a new term:
\be\label{Utrick}
S_0\rightarrow S=
S_0- \frac{3}{8\pi G }  \int d^4 x \, \phi\,  \partial_\mu T^\mu,
\ee
where for later convenience we have used:
\be
\phi=\frac{3}{\Lambda}
\ee
(the ``frequency'' appearing in the Chern-Simons-Kodama state). 
Here $T^\mu$ is a  density, so that the new term is diffeomorphism invariant without the need of a $\sqrt{-g}$ factor in the volume element or of the connection in the covariant derivative. Since the metric and connection do not appear in the new term, the Einstein equations and other standard 
field equations are left unchanged. 
The only new equations of motion are:
\begin{eqnarray}
\frac{\delta S}{\delta T ^\mu}=0&\implies& \partial_\mu \phi=\partial_\mu \Lambda =0\\
\frac{\delta S}{\delta \phi }=0&\implies& \partial_\mu  T^\mu\propto \sqrt{-g}
\end{eqnarray}
i.e. on-shell-only constancy for $\Lambda$ (the defining characteristic of unimodular theories~\cite{unimod1,unimod,alan,daughton,sorkin1,sorkin2}) and  the fact that $T^0$ is proportional to a prime candidate for relational time: 4-volume time~\cite{unimod,unimod1,UnimodLee1,Bombelli,UnimodLee2}. Reduction to minisuperspace leads therefore to:
\be
S_0\rightarrow S=
S_0 +  \frac{3 V_c}{8\pi G }  \int dt x \, \dot \phi\,  T
\ee
where $S_0$ is given by (\ref{ECaction}) and we identify $T\equiv T^0$). Classically nothing changes except that $\Lambda$'s constancy appears as an equation of motion, and the conjugate of $\Lambda$ satisfies:
\begin{equation}\label{timeformula}
\dot T =N\frac{ a^3}{\phi^2}=N\frac{\Lambda^2}{9}a^3.
\end{equation}

\subsection{Quantum unimodular theory}
However, the quantum mechanics is very different, since:
\bea\label{com2}
\left[\phi,T\right]&=&i\plk. 
\eea
Hence, we can choose either the $\phi$ (i.e. $\Lambda$) representation, leading to the standard Wheeler-DeWitt equation with solution (\ref{CS}), or the dual time representation, leading to (see~\cite{sign0,sign,CSunimod}):
\be\label{WDWSchro1}
\left[-i\plk \frac{1}{b^2+k}\frac{\partial}{\partial b}- i \plk \frac{\partial }{\partial T}\right]\psi(b,T) =0. 
\ee
that is, Schrodinger equation:
\begin{equation}\label{sch}
    i\plk \frac{\partial \psi}{\partial T}=H_0\psi
\end{equation}
with:
\begin{equation}\label{H0}
    H_0=\frac{1}{b^2+k}a^2=-\frac{i\plk}{b^2+k}\frac{\partial}{\partial b}.
\end{equation}
From the unimodular point of view, the
Chern-Simons-Kodama state (\ref{CS}) 
is just the spatial (in the sense of non-time) factor, $\psi_s$, of a monochromatic wave, with the general solution being the superposition:
\bea\label{gensolL}
\psi(b,T)&=&
\int^\infty_{-\infty} \frac{d\phi}{\sqrt{2\pi\plk}} {\cal A}(\phi) \exp{\left[-\frac{i}{\plk } \phi T \right]}\psi_s(b,\phi),
\nn\\
&=&\int^\infty_{-\infty}  \frac{d\phi}{\sqrt{2\pi\plk}} {\cal A}(\phi) \exp{\left[\frac{i}{\plk } \phi (X (b)-T ) \right]},
\eea
where the linearizing variable for the waves~\cite{sign}:
\bea
X&=&X_{CS}\equiv \frac{b^3}{3}+k b
\eea
is the Chern-Simons functional. Such wave packets can then be translated into the metric representation~\cite{HHpackets} via (\ref{FT}). For Hartle-Hawking packets (real $b$) we can also find the general solution by writing (\ref{sch}) as:
\begin{equation}\label{Sch1}
    \left(\frac{\partial}{\partial T} +  \frac{\partial}{\partial X} \right)\psi =0
\end{equation}
with solutions
\begin{equation}
    \psi(b,T)=F(T-X),
\end{equation}
explaining why $X$ is called the linearizing variable (it removes the dispersive nature of the medium).

What follows applies strictly to the Hartle-Hawking wave function~\cite{sign,CSunimod}. Then,  $b$ (and so $X$) are real and cover the whole real line. This suggests the natural inner product involving the amplitudes: 
\be\label{innalpha}
\langle\psi_1|\psi_2  \rangle=\int^\infty_{-\infty} d\phi\,    {\cal A}_1^\star (\phi) {\cal A}_2(\phi) .
\ee
Given the Fourier form in the variable $X$ of the superpositions (\ref{gensolL}), we can use the inverse Fourier transform to find:
\begin{equation}\label{inverseFTX}
    {\cal A}(\phi)  e ^{-\frac{i}{\plk } \phi T }=
\int^\infty_{-\infty}  \frac{d X}{\sqrt{2\pi\plk}} \psi(b(X),T) e^{-\frac{i}{\plk } \phi X }
\end{equation}
assuming that $X$ and $\phi$ vary without any constraints on the real line (this is to be seen as the boundary condition, or lack thereof, for the imposition of unitarity). This FT is not to be confused with (\ref{FT}) relating metric and connection representations. This is the advantage of using the connection representation: instead of Airy functions, the $\psi_s$ are {\it plane} waves in the  Chern-Simons functional. 

We can therefore identify an equivalent inner product to (\ref{innalpha}) in the connection representation. Either by using (\ref{inverseFTX}) or by invoking 
Parseval's theorem, Eq.~(\ref{innalpha}) is equivalent to:
\be\label{innX}
\langle\psi_1|\psi_2  \rangle=\int^\infty_{-\infty} dX  \psi_1^\star(b,T)\psi_2(b,T)
\ee
which is time-independent (since it is equal to (\ref{innalpha}) where $T$ does not feature), so that unitarity has been enforced. Hence the probability in term of $b$ is:
\be
{\cal P}(b)=|\psi(b,T)|^2\left|\frac{dX}{db}\right|
\ee
where we stress a measure factor multiplying  the Born $|\psi|^2$: the derivative of the Chern-Simons functional (and so the corresponding terms in the Hamiltonian constraint).

With these definitions we have unitary evolution. We also recover the classical limit (given by $\dot X=\dot T$) from the peak of the probability for Gaussian states (see~\cite{sign,HHpackets} for details). 

\section{Failure of the unimodular approach in Vilenkin's case}\label{VLap}

Regrettably, this approach---applicable to the Hartle-Hawking wave function---is lost in translation into the V wave function. This happens because $b$ has an imaginary section~\cite{CSHHV} in order to reproduce the V wave function, and therefore so does the Chern-Simons functional $X$. Thus, the assumptions for using the FT break down.

We could take the view that Hartle-Hawking and Vilenkin are analytical continuations of each other (something pointed out at least as far back as in~\cite{vil-PRD}), so that we should replace the FT transform by the complex Laplace transform to accommodate the V wave function. The unimodular extension then still leads to the superpositions  (\ref{gensolL}) seen as 
the Laplace transform:
\be\label{LT}
F(s)=\int f(t)e^{-st} dt, 
\ee
with real $t$ and complex $s$, after identifications: 
\bea
t&=&\phi\nn\\
f&=&{\cal A}(\phi)e^{-\frac{i}{\plk}\phi T}\nn\\
s&=&-\frac{i}{\plk} X.
\eea
(up to conventional factors of $2\pi$). What follows applies both for $t\in (-\infty,\infty))$ leading to the bilateral Laplace transform, or for  $t\in (0,\infty))$, leading to the standard Laplace transform\footnote{It may seem that this makes a difference for convergence. Usually ${\cal A}$ is a  Gaussian centered on a positive $\phi_0$: this has a tiny support in $\phi<0$. However, for the V contour it would make the wave packets divergent as $b\rightarrow -i\infty$. As, it happens, the problems are more serious, and restricting $\phi>0$ makes no difference.}.

The problem appears in the inversion formula (\ref{inverseFTX}) and consequent use of Parseval's theorem (both of which have versions for either the standard or the bilateral Laplace transform). The inversion formula is:
\be
f(t)=\frac{1}{2\pi i}\int ^{\gamma +i\infty}_{\gamma -i \infty} e^{st}F(s)\,  ds
\ee
where $\gamma$ is a real number so that the   contour is in the region of convergence of $F(s)$. We can always choose $\gamma=0$ so that the contour is the Hartle-Hawking contour. Consequently Parseval's theorem, leading to the probability interpretation, reads:
\bea
\langle\psi_1|\psi_2  \rangle
&=&\int d\phi  {\cal A}_1^\star(\phi) {\cal A}_2(\phi) = \int_{-\infty}^\infty dX  \psi_1^\star(b,T)\psi_2(b,T).\nn
\eea
The relevant integrations leading to the amplitudes and the probability density only care about the wave function along the Hartle-Hawking contour. 

This is a damning conclusion. If we see Hartle-Hawking and Vilenkin as analytically related, and apply the unimodular prescription to define a conserved inner product, we find that the inner product expressed in terms of the connection  ignores the analytical extension leading to V, and refers us back to the Hartle-Hawking theory as if no analytical extension had been made.



\section{The nail in the coffin}\label{nail}
This is unsurprising. In Section~\ref{HH} we derived unitarity {\it directly} from the fact that the inner product is defined via Eq.~(\ref{innalpha}) in terms of $T$-independent amplitudes. We then derived the  equivalent expression (\ref{innX}), in terms of the time-dependent wave functions, which must therefore also be $T$ independent. The usual way to prove unitarity, via the Hermiticity of the Hamiltonian, can therefore be bypassed, but nonetheless it sheds light on why the argument in 
Section~\ref{HH} fails for the Vilenkin contour.

Indeed we could have integrated (\ref{sch}) via the evolution operator (with respect to $T$):
\begin{equation}
    U=\exp\left[-\frac{i}{\plk}\int dT H_0\right].
\end{equation}
The conservation of $\langle\psi_1|\psi_2\rangle $ is enforced by $U^\dagger=U^{-1}$, i.e.  the unitarity of $U$ (and hence the terminology). This is guaranteed by the Hermiticity of $H_0$ ($H_0^\dagger=H_0$) which is straightforward to prove under inner product (\ref{innX}), assuming $X$ is unbounded and real.

For the Hartle-Hawking wave function this is a very roundabout way to prove the time-independence of an inner product which is true by construction; but it shows why the equivalent construction for the Vilenkin wave function (allowing for a imaginary $b$ and $X$) cannot work. The problem is fundamentally that 
$H_0$ is anti-Hermitian if $b$ is imaginary. 

Another way to see how this comes about concerns the last identity in (\ref{H0}),  arising from:
\begin{equation}
     [b,a^2]=i \plk\implies a^2=-i\plk \frac{\partial}{\partial b},
\end{equation}
implying that the Hamiltonian with inner product (\ref{innX}) is anti-Hermitian whenever $b$ is imaginary. This should not be confused with the fact that $a^2$ is not Hermitian even for the Hartle-Hawking contour, where $b$ is real. Note that $[b,a^2]=i \plk $ implies:
\begin{equation}
     [b^\dagger,(a^2)^\dagger]=i \plk
\end{equation}
but for real (Hermitian) $b$ the fact that $a^2$ is non-Hermitian does not conflict with this because:
\begin{equation}
    (a^2)^\dagger=a^2+\frac{2i\plk b}{b^2+k}.
\end{equation}
However, if $b$ has non-real eingenvalues then it also cannot be Hermitian. Where $b$ is imaginary, the first term in $(a^2)^\dagger$ is $-a^2$, instead of $a^2$, creating the real problem.

Finally, we note that there are other trivial constructions for Hartle-Hawking that do not carry over to the Vilenkin setting. 
For example, for real $T$ and $X$ we can write the solutions to (\ref{Sch1})
as $\psi(b,T)=F(T-X)$.
This lies behind the conservation of probability with measure $dX$ (the wave functions are non-dispersive  in $X$). If $X$ is not real this argument collapses. 



\section{Euclidean image in Lorentzian theory}\label{EucImage}

Had we started from an Euclidean signature (with all quantities taking real values), only the sign of the kinetic term, $b^2$, in action (\ref{ECaction}) would be modified:
\begin{equation}\label{Sg}
S_E=\frac{3V_c}{8\pi G} \int 
dt\bigg(\dot{b}a^2-Na \left[ b^2-k 
 + \frac{a^2}{\phi}\right]\bigg).
\end{equation}
This can be worked out from first principles, but is often inferred from an analytical extension of the Lorentzian theory.
Writing:
\begin{eqnarray}
    t&= &\pm i\tau
\end{eqnarray}
with fixed lapse function $N$  (or equivalently $N=\pm i\tilde N$ with coordinate $t$ fixed)\footnote{The sign ambiguity is a source of great controversy, but it will play no role in what follows.}
we find that the Lorentzian action (\ref{ECaction}) becomes:
\begin{equation}
    S_L=\pm i \frac{3V_c}{8\pi G }\int d\tau \bigg( \bar b'a^2-Na\left[\bar b^2-k+\frac{a^2}{\phi}\right]\bigg)
\end{equation}
with $'=d/d\tau$ and:
\begin{eqnarray}
     b=\pm i\bar b.
\end{eqnarray}
Comparing with (\ref{Sg}) we see that 
the imaginary Lorentzian action
is related to their the Euclidean counterpart by: 
\begin{eqnarray}
  S_L[b,a^2,t]&=&\pm iS_E[\bar b,a^2,\tau]. 
\end{eqnarray}
We can also consider an Euclidean version of the unimodular extension mimicking:
\begin{eqnarray}
   S_E&\rightarrow & S_E+\frac{3V_c}{8\pi G}\int d\tau \phi'\bar T \\
  T&=&i\bar T ,
\end{eqnarray}
so Euclidean unimodular time can be seen as imaginary Lorentzian time and vice versa.  Finally, we can define an Euclidean version of the Chern-Simons functional: 
\begin{eqnarray}
   X(b)&=&i\bar X(\bar b) \\
\bar X&=&-\frac{\bb^3}{3}+k\bb. \label{CSeuc0}
\end{eqnarray}



Within a Lorentzian theory allowing the connection $b$ to become imaginary is equivalent to an Euclidean theory with a real connection $b$; however the Lorentzian action is now  imaginary.
Therefore, the Vilenkin contour $b=b\Theta (b)+i\bar b\Theta(-\bar b)$ implies that the action is of the form: 
\begin{equation}\label{STheta}
    S[b]=\Theta(b)S_L\pm i \Theta(-\bar b) S_E[\bar b].
\end{equation}
The excursion of $b$ into the imaginary domain can be seen as {\it an Euclidean image within the Lorentzian theory}, that is, keeping the fundamental action as $S_L$, so that the action is imaginary in the relevant domain. This is the origin of all the problems with unitarity.

Wherever the action is imaginary, the Hamiltonian is imaginary. The Poisson brackets are also imaginary, so that:
\begin{eqnarray}
    \left[\bar b,a^2\right]&=&\plk\\
    \left[\phi,\bar T\right]&=&\plk
\end{eqnarray}
for the Hermitian $\bar b$ and $\bar  T$. Upon quantization with these rules the Hamiltonian therefore is anti-Hermitian, just as we found in Section~\ref{nail}. The wave functions in connection space, along the imaginary section of the connection, are real and evanescent. Unitarity is lost. 

The conflict between unitarity and the Vilenkin proposal is therefore deeply ingrained in the structure of the theory and it is hard to see how it can be bypassed.


\section{Euclidean action and signature change}\label{EucSignChange}
In the above it is easy to forget that the Euclidean theory exists in its own right, with a real action and connection, without reference to the Lorentzian theory. We could have associated the Vilenkin contour with a different theory, in which there is an actual signature change at $b=0$, from Lorentzian to Euclidean, the connection remaining real. Instead of (\ref{STheta}) the action then is:
\begin{equation}\label{Sgs}
    S_0[b]=\Theta(b)S_L+ \Theta(- b) S_E[ b],
\end{equation}
that is, the perennially real action:
\begin{equation}
S_0=\frac{3V_c}{8\pi G}\int{\rm d}t\left(\dot{b}a^2-Na\left(s(b)b^2-k+\frac{a^2}{\phi}\right)\right)
\end{equation}
where 
\begin{equation}
    s(b)=-\Theta(b)+\Theta(-b). 
\end{equation}
The commutators are now imaginary and the Hamiltonian is Hermitian, as they should be. 
Since the action is always real, there is no need to complexify the time variable in the unimodular extension in the Euclidean section:
\begin{eqnarray}
   S &\rightarrow&S+\frac{3V_c}{8\pi G}\int d t \dot \phi T. 
\end{eqnarray}
Unimodular time remains real throughout.

The Hamiltonian constraint and EOM for this theory are:
\begin{eqnarray}
0&=& s b^2-k+\frac{a^2}{\phi} \\
   \dot a&=&-s N b -Nb^2 \delta(b)=-s N b \label{dota}\\
  \dot b &=&N\frac{a}{\phi}\\
   \dot  T&=&N\frac{a^3}{\phi^2}.
\end{eqnarray}
Since the change in $s$ happens at $b=0$ no new term appears in (\ref{dota}). The usual consistency relations obtained from dotting the Hamitonian constraint and comparing with the EOM are satisfied, i.e. there is no conflict with the Bianchi identities.

The classical solutions are a conflation of the Lorentzian and Euclidean solutions. For $t>0$ we have:
\begin{eqnarray}
   a(t)&=&\sqrt{k\phi}\cosh (t/\sqrt{\phi}) \label{aE0}\\
  b (t) &=&\sqrt{k} \sinh (t/\sqrt{\phi}) .\label{bE0}
\end{eqnarray}
and for $-\pi\sqrt{\phi}/2<t<0$:
\begin{eqnarray}
   a(t)&=&\sqrt{k\phi}\cos(t/\sqrt{\phi}) \label{aE}\\
   b (t) &=&\sqrt{k} \sin (t/\sqrt{\phi}) .\label{bE}
\end{eqnarray}
corresponding to the iconic half deSitter glued to half a 4-sphere, except that this now is a classical solution. The Chern-Simons functional of this theory is always imaginary (so the wave functions are never evanescent; they propagate even in the Euclidean phase, as we shall see), that is $X$ is always real with:
\begin{eqnarray}\label{Xs}
    X=\int db\,  (-sb^2+1)=-s\frac{b^3}{3}+k b
\end{eqnarray}
matching (\ref{CSeuc0}) for $b<0$. The unimodular time $T$ is also real and given by:
\begin{equation}
    T=\int dt\, N\frac{a^3}{\phi^2}.
\end{equation}
It is easy to find its on-shell expressions.  
For example for $t<0$ we have
 \begin{eqnarray}
    T&=&\int dt\, \frac{a^3}{\phi^2} \nn \\
    &=& k^{3/2}\left[-\frac{3}{4}\cos\left(\frac{ t }{\sqrt{\phi}}\right)+\frac{1}{12}\cos\left(\frac{3 t }{\sqrt{\phi}}\right)\right]\nn \\
    &=&k^{3/2}\left[-\frac{1}{3}\sin^3\left(\frac{ t }{\sqrt{\phi}}\right)+ \sin\left(\frac{t }{\sqrt{\phi}}\right)\right]
\end{eqnarray}
where we have adjusted the integration constant in $T$ so that  $T=0$ marks the transition from Euclidean to Lorentzian signature. 
The solutions satisfy:
\begin{eqnarray}
    X=T
\end{eqnarray}
throughout. No issues should therefore arise with unitarity on the grounds highlighted in the previous Section. Conserved (non-dispersive in $X$) travelling waves can be expected in the quantum theory.

However we note a novelty. The point $b_\emptyset=-\sqrt{k}$ is now the endpoint of the manifold, or the no-boundary South Pole. Hence the Chern-Simons functional $X$ and unimodular time $T$ are limited from below by:
\begin{equation}
    T_\emptyset=-\frac{2}{3}k^{3/2}
\end{equation}
marking the beginning of the Universe.
Far from being a no-boundary, unimodular time has a boundary. 
Therefore, we are still not off the hook regarding unitarity, since we have a theory with a boundary. We now proceed to make this problem explicit and solve it. 




\section{The Euclidean quantum theory}\label{EucQuantum}
Let us first study unitarity in an Euclidean theory with real action (rather than its imaginary image as seen from the Lorentzian theory). Since the theory's variables and Hamiltonian are real, 
a priori there should be no problems quantizing it to obtain a  unitary theory. 
Eq.~(\ref{Sgs}) with $s=1$ implies:
\begin{eqnarray}
\{b,a^2\}&=&\frac{3V_c}{8\pi G}\\
H_0&=&\frac{1}{-b^2+k}a^2
\end{eqnarray}
(where $b,a^2\in \mathrm{R}$), and quantization proceeds as:
\begin{eqnarray}
   [b,a^2]=i \plk&\implies& a^2=-i\plk \frac{\partial}{\partial b}\nn \\ 
    H_0&=&-\frac{i\plk}{-b^2+k}\frac{\partial}{\partial b}=-i\plk \frac{\partial}{\partial X}
\end{eqnarray}
where
\be\label{XEuc}
X=-\frac{b^3}{3}+k b
\ee
is the (real) Chern-Simons functional adapted to Euclidean signature. 
We therefore recover Schr\"odinger equation (\ref{sch}) with a $H_0$ which is Hermitian with inner product (\ref{innX}) if $b$ and $X$ are allowed unrestricted variation. There are no obstructions to rewriting this as (\ref{Sch1}) with this $X$, with solutions $\psi(b,T)=F(T-X)$. 
It is straightforward to see that if $X$ and $T$ are unrestricted, the methods in~\cite{HHpackets} for enforcing unitarity follow through. 

\begin{figure}
\centering
\includegraphics[scale=0.53]{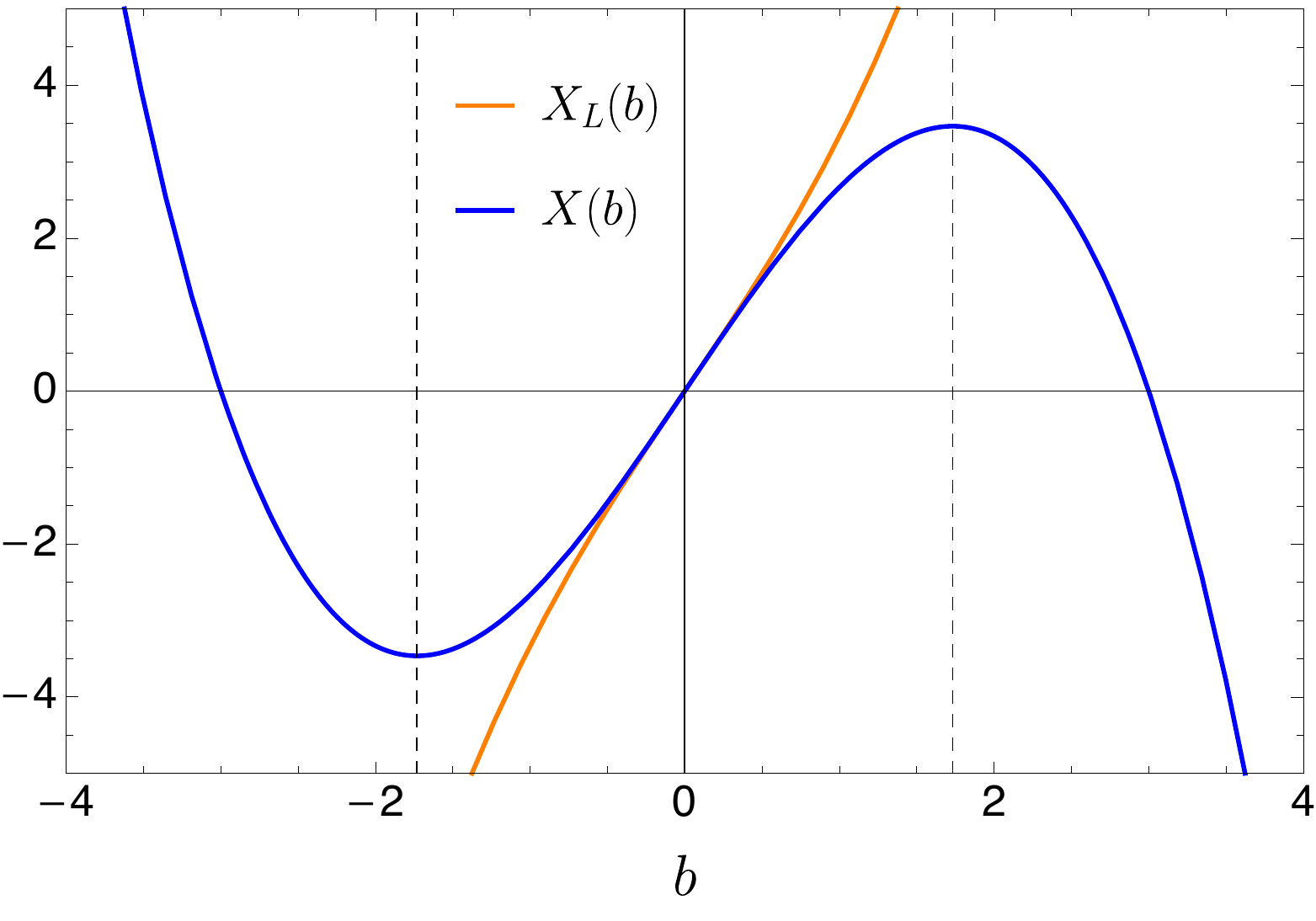}
\caption{Chern-Simons function for the Lorentzian (L) and Euclidean cases in terms of $b$ with $k=1$. The latter is non-monotonic if we do not restrict $b\in (b_-,b_+)$, that is, to the classically allowed region (from North to South pole of the classical 4-sphere). This restriction is indicated by the dashed lines. } 
\label{figX}
\end{figure}

However, unrestricted variation in $b$ and $X$ in this context is problematic. 
The catch is in that the Euclidean Chern-Simons function (\ref{XEuc}) upon which wave functions depend (and providing the inner product measure) is non-monotonic, so that the inverse $b(X)$ is multi-valued if we do not restrict to $b\in (-\sqrt {k},\sqrt{k})$ (see Fig.\ref{figX}). The projection of the norm (\ref{innX}) onto $b$ is ill-defined if no restrictions are applied. 
We should therefore restrict the $b$ domain as:
\be
b\in (b_-,b_+); \quad b_\pm=\pm\sqrt{k}
\ee
and so:
\be
X\in (X_-,X_+); \quad X_\pm=X(b_\pm)=\pm \frac{2}{3} k^{3/2}.
\ee
This makes sense from the point of view of the classical theory, since the 
classical solution is a 4 sphere, which can be parameterized by (\ref{aE}) and (\ref{bE}), implying $b\in (-\sqrt {k},\sqrt{k})$. 
However, 
the quantum theory does not {\it a priori} need to mimic the classical solution: the wave function could be an evanescent wave outside $b\in (-\sqrt {k},\sqrt{k})$. But as we have seen, the reason for excluding this is that $b(X)$ would then be multivalued, rendering $\psi(b)$ undefined. 

Imposing this restriction
leaves us in territory familiar to the literature using the metric formalism~\cite{GielenMenendez}: if variables are restricted, then the Hamiltonian is Hermitian only if suitable boundary conditions are imposed.  The standard argument applied to our setup reads:
\begin{eqnarray}
     \langle \psi_1|H|\psi_2\rangle&=&\int^{X_+}_{X_-} dX
    \psi_1^\star (-i\plk) \frac{\partial}{\partial X}
    \psi_2\nn\\
  &=&   \langle \psi_2|H|\psi_1\rangle^\star-i\plk [ \psi_1^\star  \psi_2 ]^{X_+}_{X_-}\label{Hherm}
\end{eqnarray}
i.e. the integration by parts generates two boundary terms which require more onerous boundary conditions than the simple fall off conditions for when the limits lie at infinity.

Discounting non-local boundary conditions (e.g. $\psi(b_-,T)=\psi(b_+,T)$, transferring the wave function non-locally from North to South pole of the {\it same} sphere), the obvious solution is 
reflecting boundary conditions:
\begin{equation}\label{reflectBC}
    \psi(b_\pm, T)=0
\end{equation}
(where $T$ is real and unrestricted). However this creates a number of technical problems (as we explain briefly in Section~\ref{other}). Also, it has the implication of reflecting the wave back into the Lorentzian domain. We then find a contracting and an expanding Universe in the Lorentzian region, in contradiction with Vilenkin's proposal, where only an expanding travelling wave should be present. 

\section{Sisyphus boundary conditions}
\label{Sisyphus}
One alternative, leading to an outgoing but no incoming wave in the Lorentzian region, is a semi-infinite tower of Euclidean spheres. To Hawking's rhetorical question, ``what is south of the South pole?'', one can retort (as did~\cite{Tom1,Tom2}): ``another North pole". This ``south of nowhere''  process can be continued ad (semi) aeternum, from South of the $n$  sphere to North of the $n-1$ sphere; indeed  this is required should  we insist on unitarity together with an outgoing without incoming wave in the Lorentzian region. A reflection at any South pole down the tower of spheres would only increase the interval in unimodular time between the incoming and  outgoing waves in the Lorentzian region\footnote{Also, to be pedantic, going backwards in unimodular time, any reflection on a North pole after a reflection off a South pole would only send the wave down the tower, ultimately visiting the semi-infinite tower of spheres.}. 

\begin{figure}
    \centering
\includegraphics[scale=0.53]{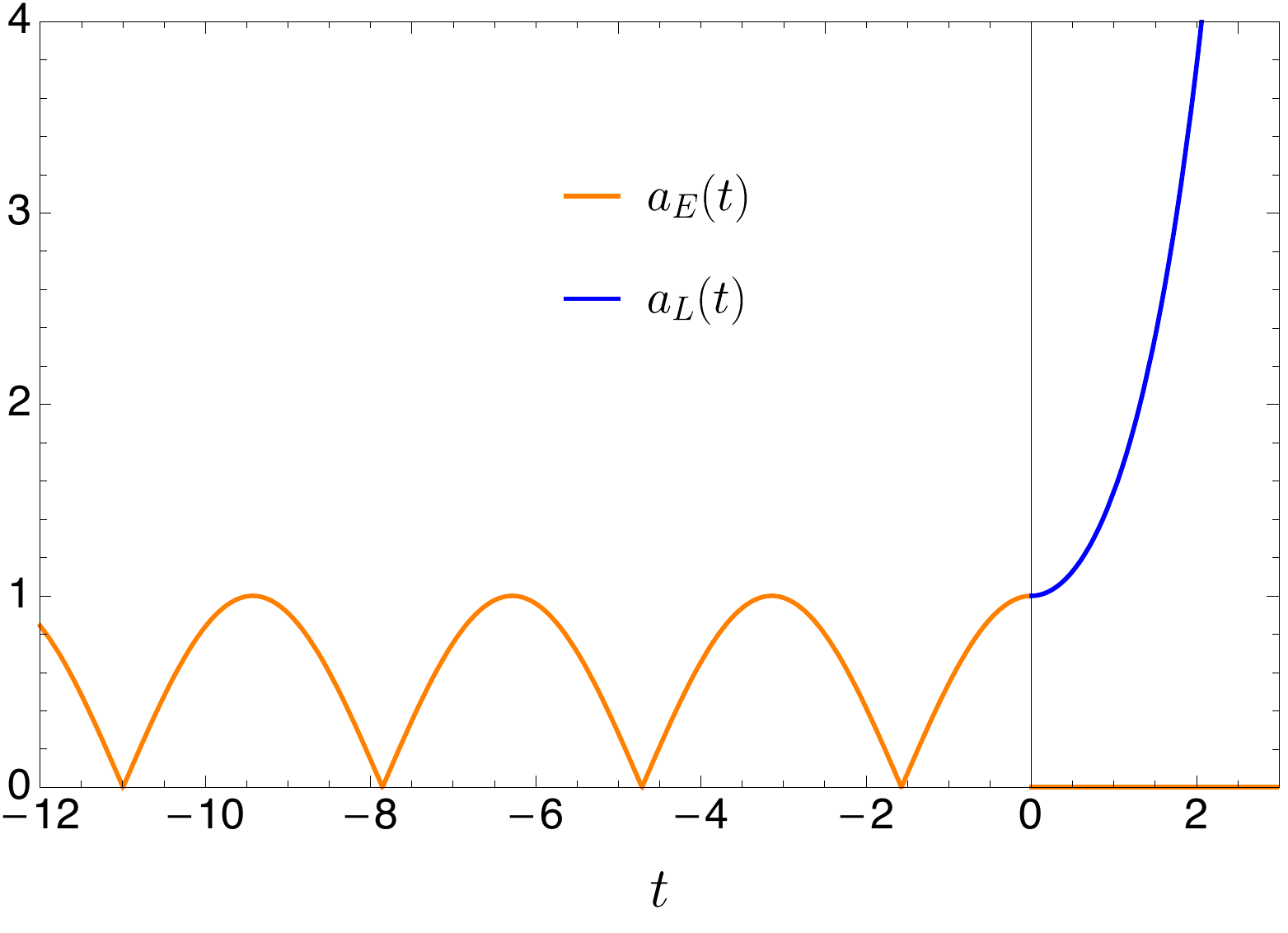}
\includegraphics[scale=0.53]{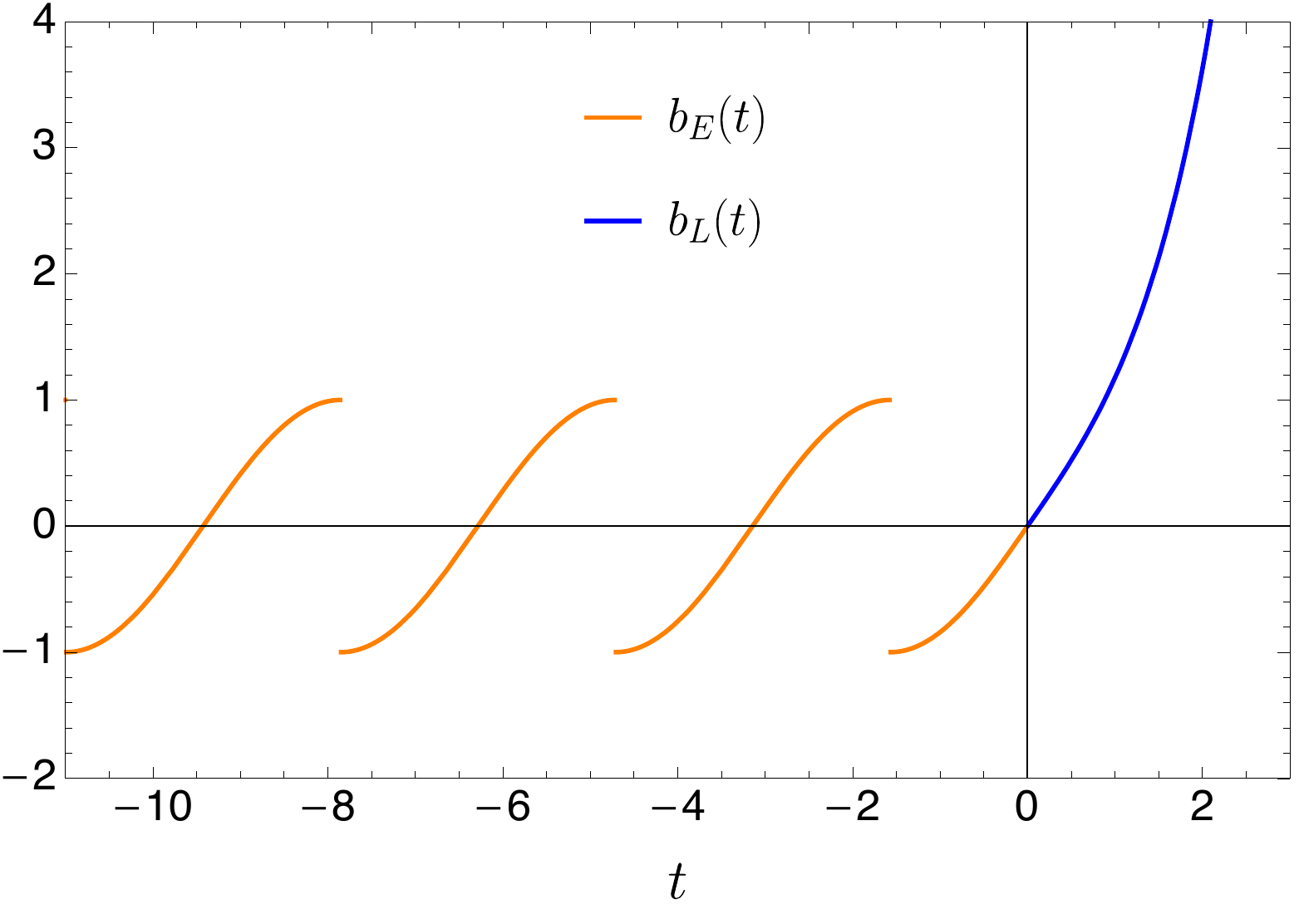}
\caption{Top panel: scale factor $a$ for the Euclidean (subscript E) and Lorentzian (L) cases as a function of coordinate time $t$, allowing for a $-\infty<n\le 0$ tower of spheres. Bottom panel: connection variable $b$ for the Euclidean (E) and Lorentzian (L) cases as a function of coordinate time $t$ in the same setup. In both cases we have considered $k=1=\phi$. } 
\label{figab}
\end{figure} 

\begin{figure}
\centering
\includegraphics[scale=0.5]{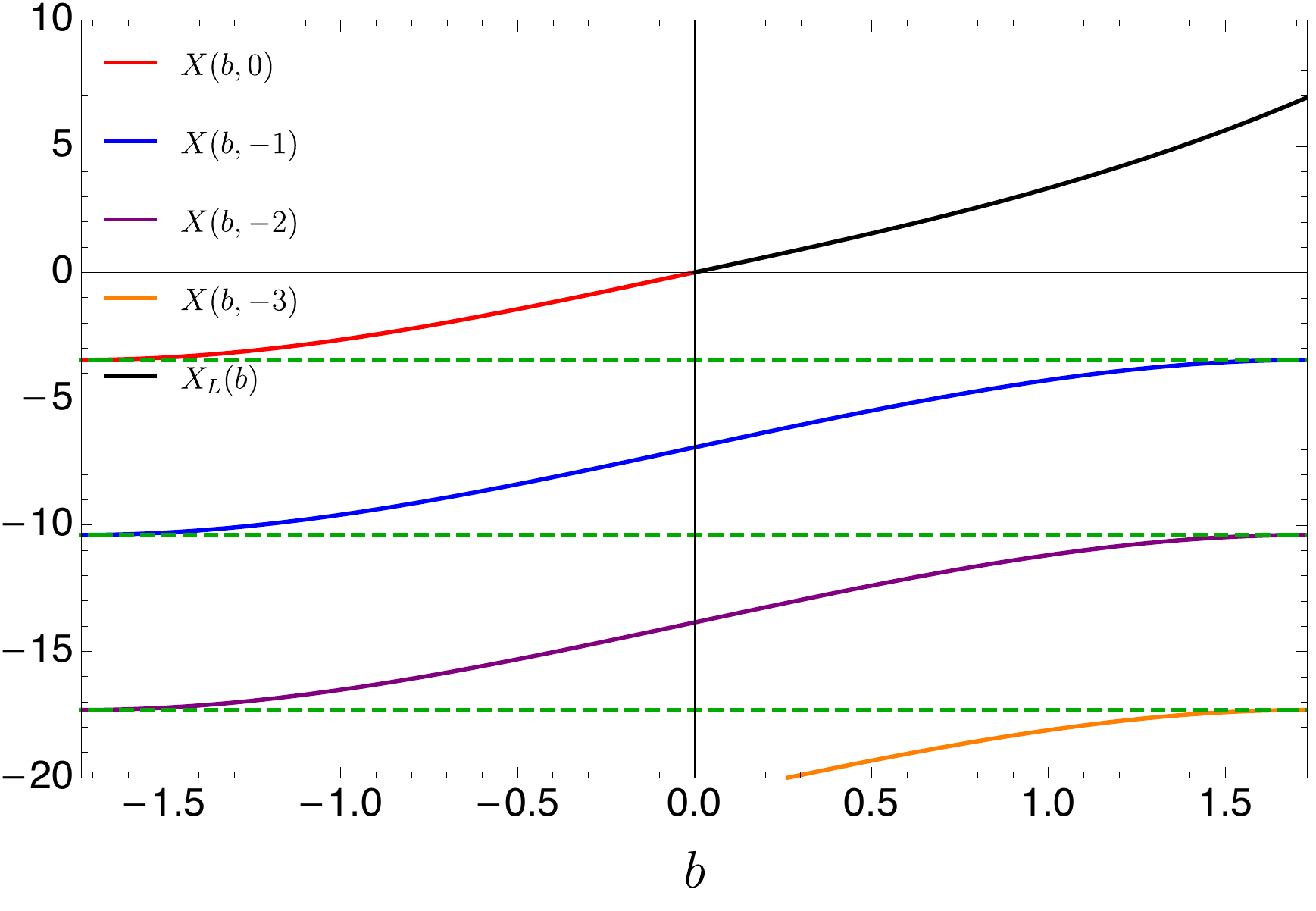}
\caption{The linearizing variable $X$ solving the Schrodinger equation, connecting the Lorentzian outgoing $X$ (top right) with the the Euclidean $X(b,0)$ at $b=0$, followed  by a semi-infinite Euclidean tower, represented by  $X(b,n)$
with $-1\le n<\-\infty$. Notice how the South pole of the $n$ sphere and North pole of the $n-1$ sphere have the same $X$.  }
\label{figXn}
\end{figure}

 \begin{widetext}
  \end{widetext}  
    \begin{figure}
\centering
\setlength{\tabcolsep}{0.1pt}
\begin{tabular}{c c}
\includegraphics[scale=0.575]{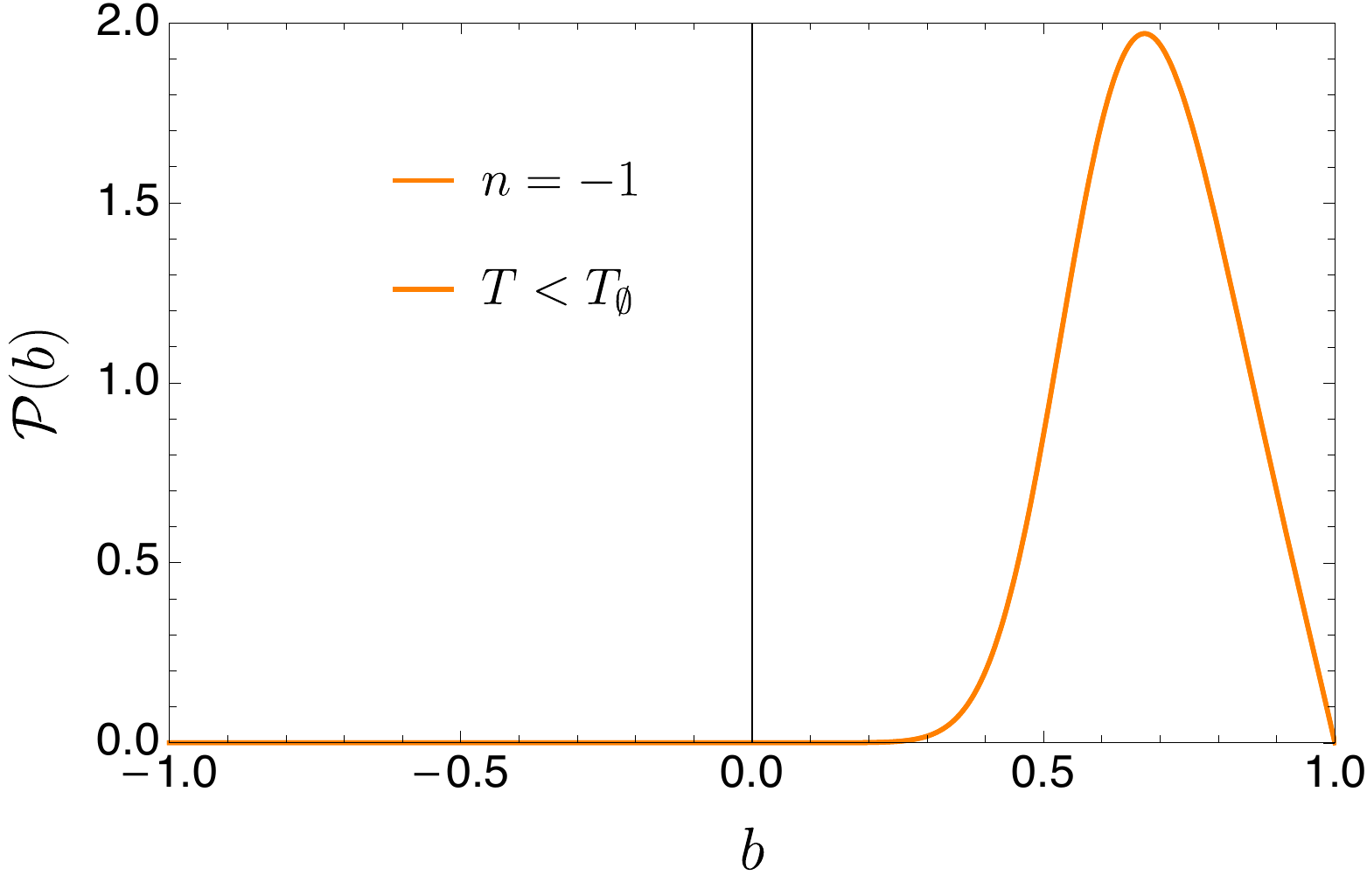} & \includegraphics[scale=0.53]{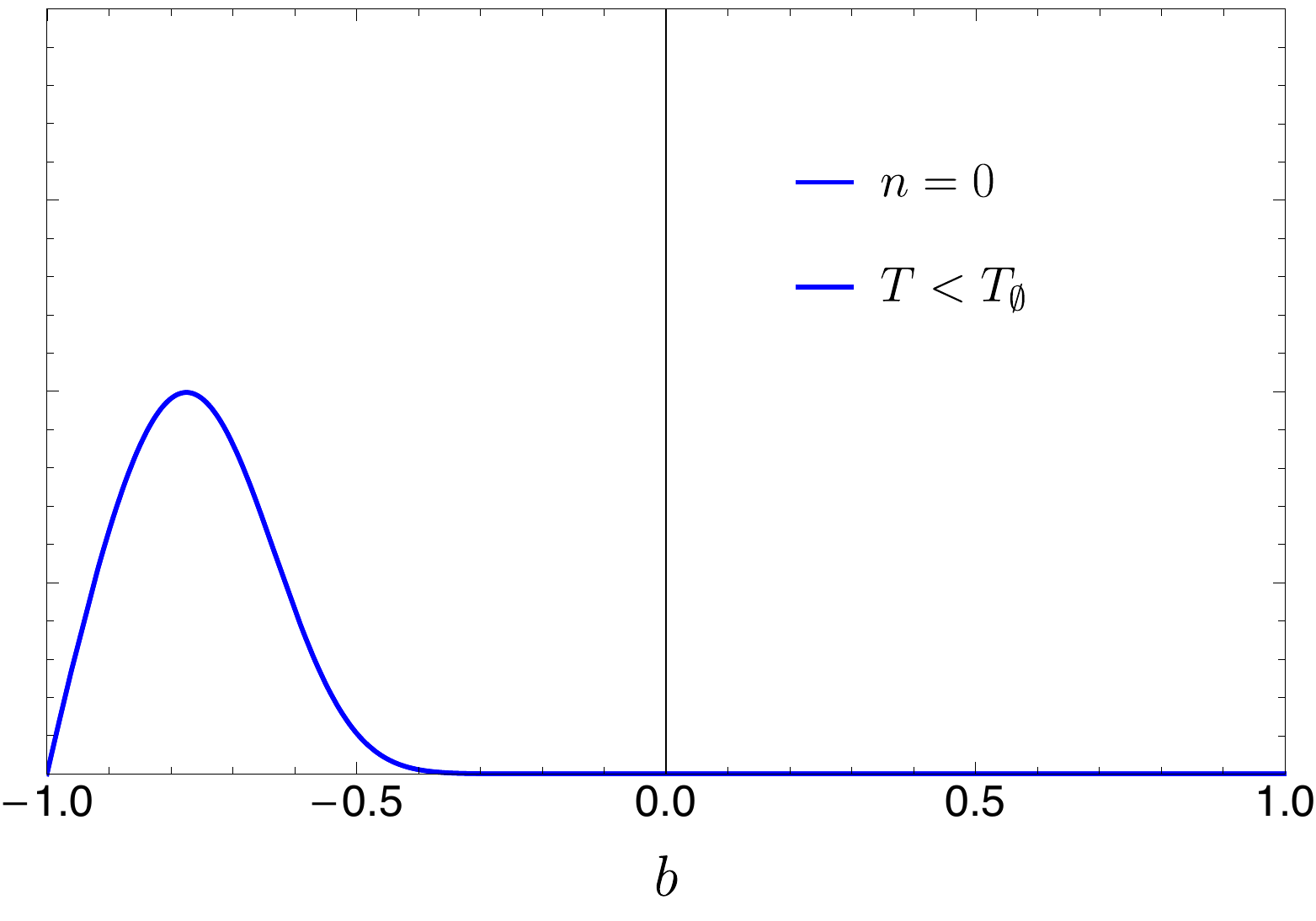} \\
\includegraphics[scale=0.575]{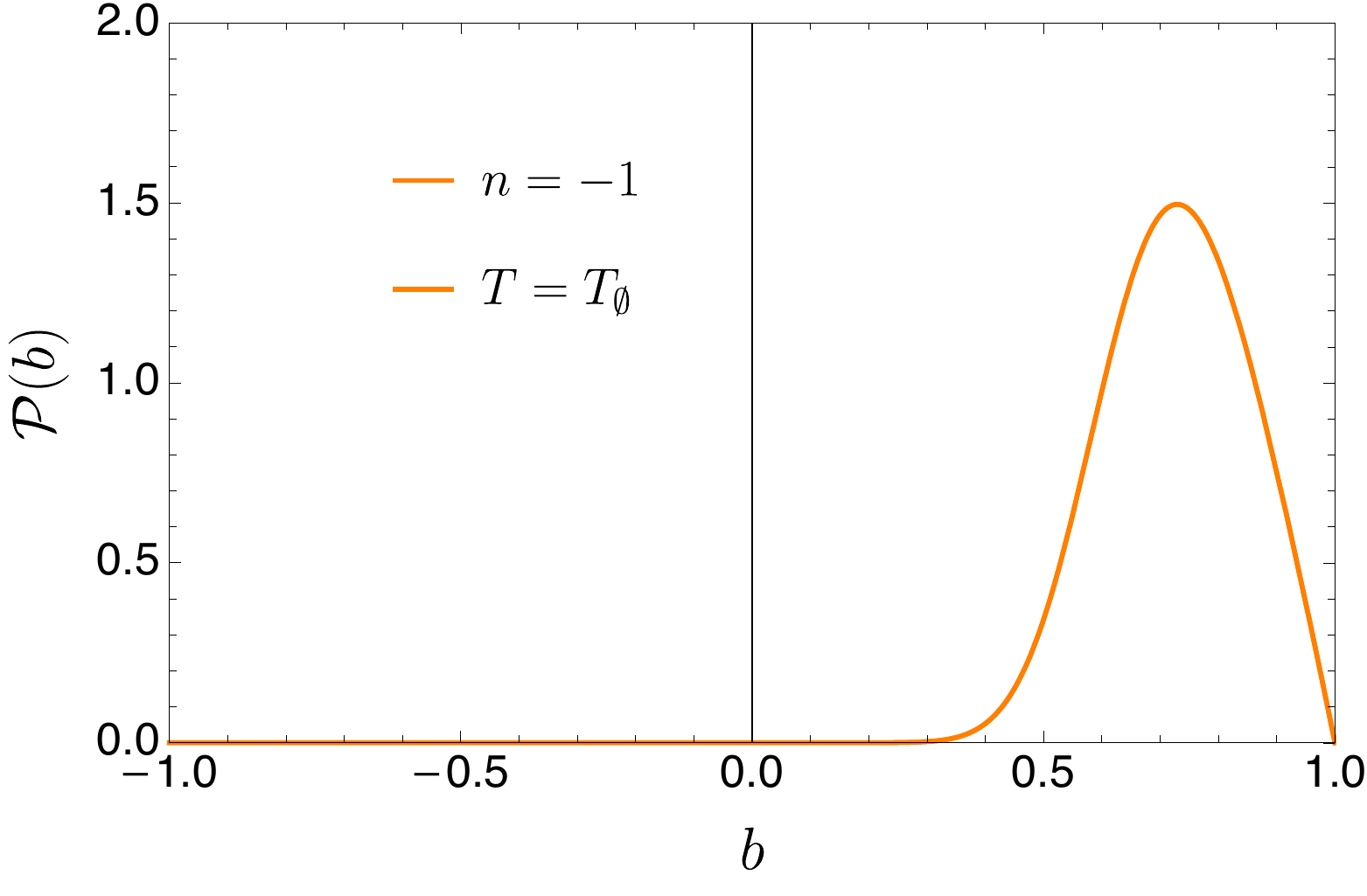} & \includegraphics[scale=0.53]{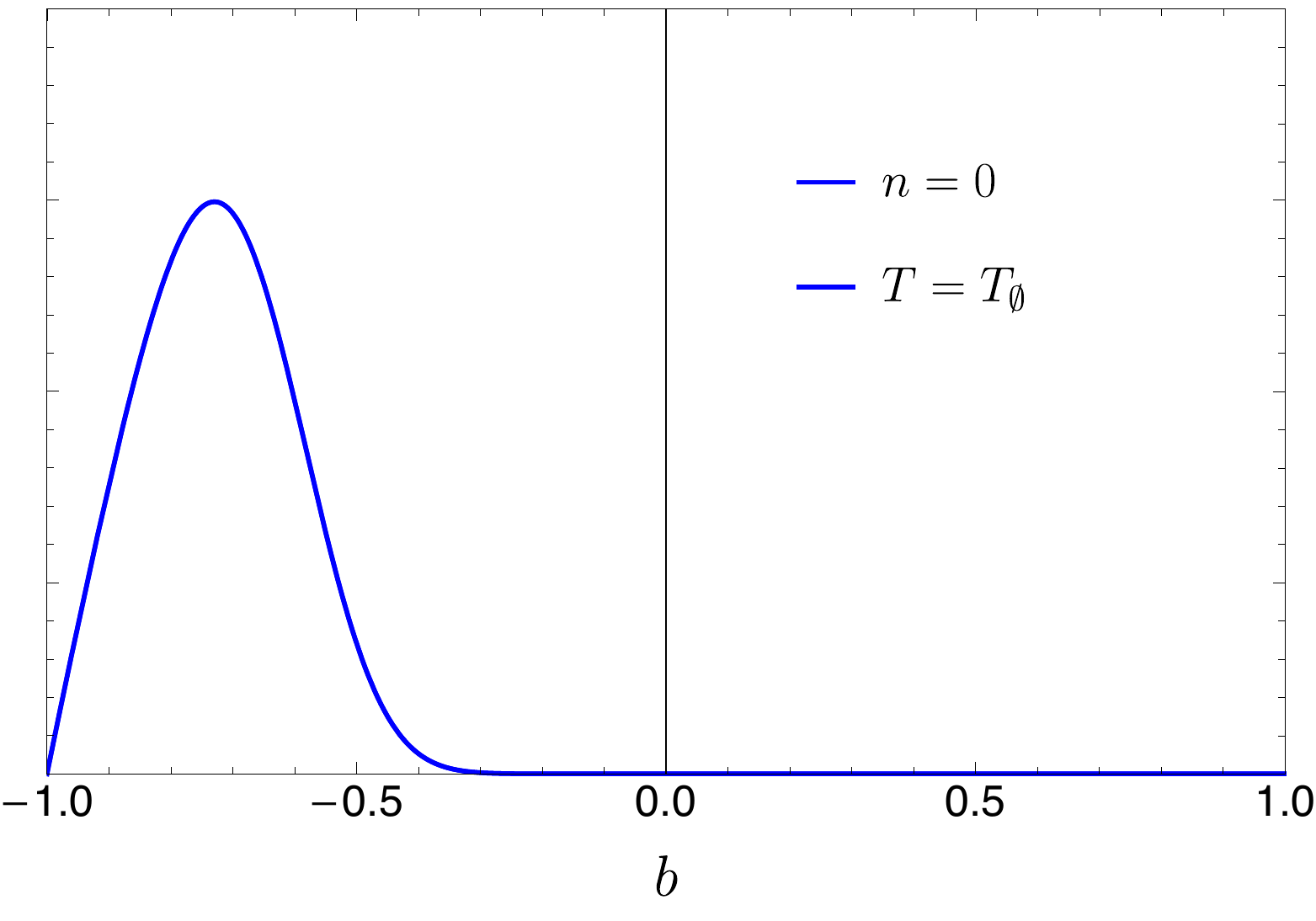} \\
\includegraphics[scale=0.575]{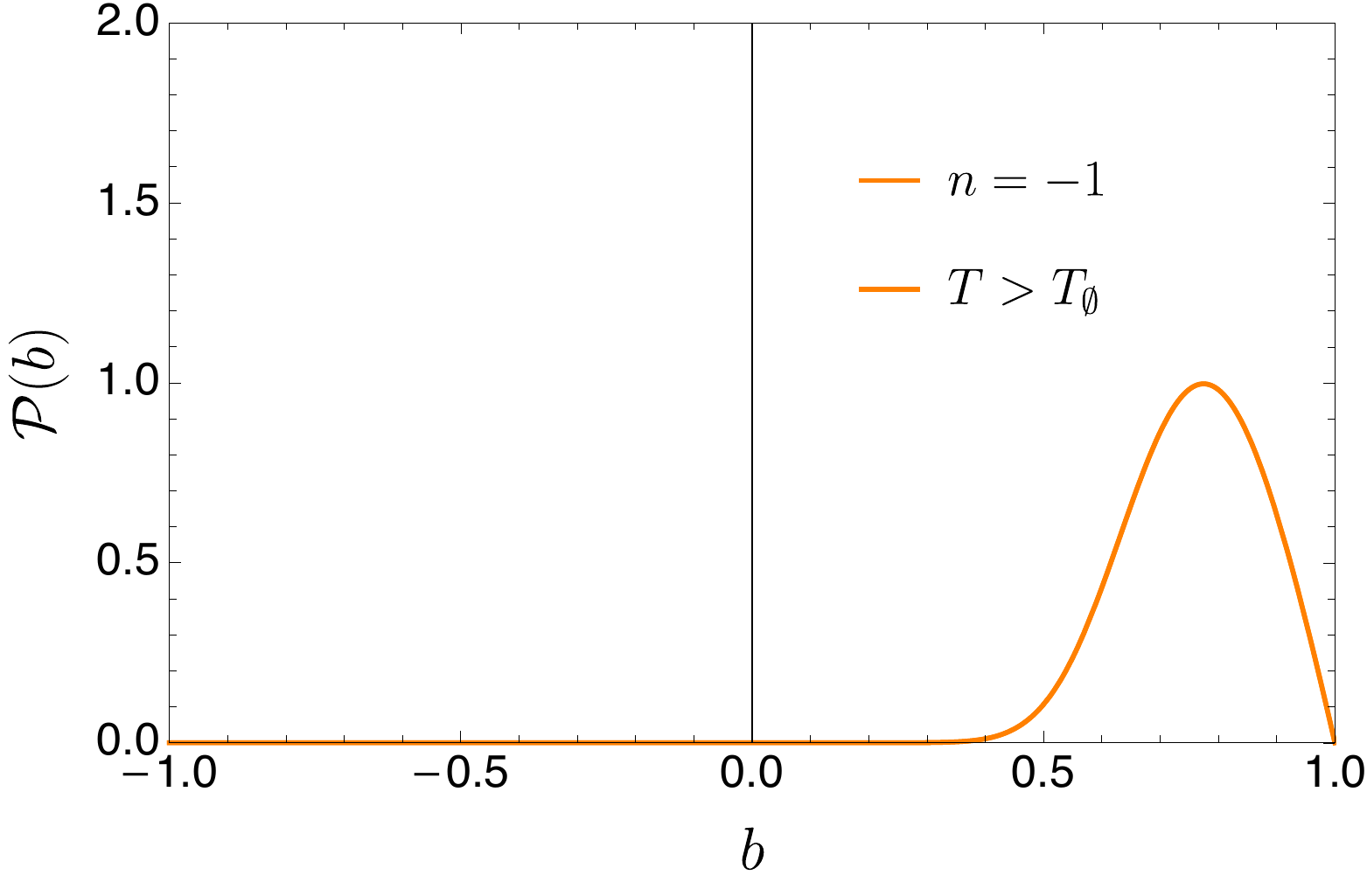} & \includegraphics[scale=0.53]{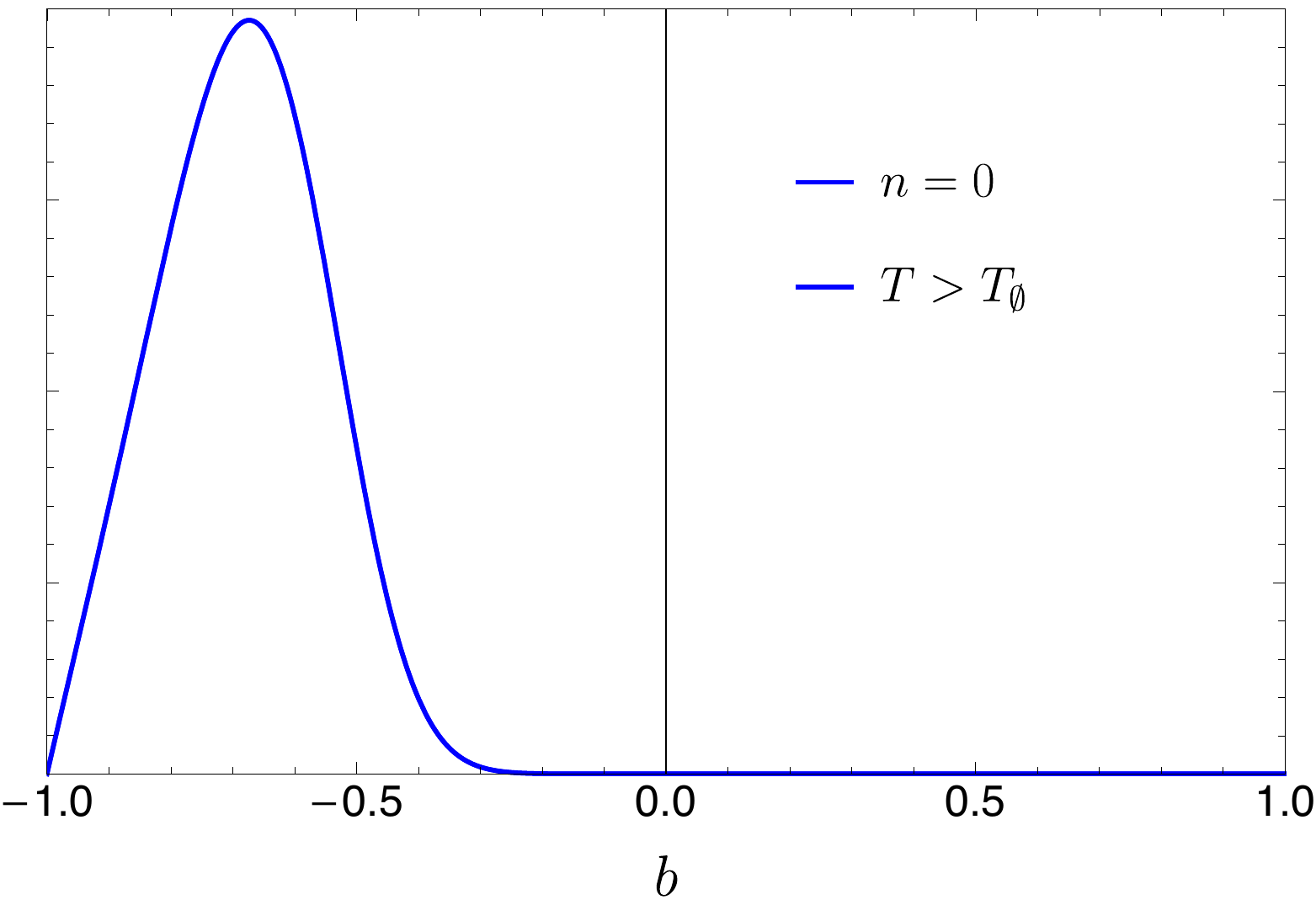} \\
\end{tabular}
\caption{Probability density as a function of $b$ 
as $T$ increases from just below to just above $T_\emptyset$, and the peak of the wave function progresses from just below the North pole of the $n=-1$ sphere to the South pole of the $n=0$ sphere. As we see, the wave function departs from the classical trajectory, as this approaches the singular point where North and South meet. Instead two peaks establish themselves at the two spheres at an equidistant latitude determined by $\sigma_T$. As time progresses the peak at $n=0$ rises at the expense of the one at $n=-1$. At $T=T_\emptyset$ (or at any other time where the classical trajectory would be singular) we have a perfectly balances superposition.}
\label{figprob}
\end{figure}

An infinite tower of spheres is already in use as a mathematical device for implementing reflecting boundary conditions for a single sphere via an ``images'' method (as we will review in Section~\ref{other}). In that context only the $n=0$ sphere exists, but one avails oneself of infinite copies with identifications:
\begin{eqnarray}
    N_0&\equiv &S_{2n+1}\equiv N_{2n}\nn\label{equiv1}\\
    S_0&\equiv &N_{2n+1}\equiv S_{2n}\nn\label{equiv2}
\end{eqnarray}
to obtain the wave function satisfying the boundary conditions, as it bounces back and forth between North and South poles. In contrast, we posit that the spheres are distinct, with each South pole glued to the North pole below:
\begin{eqnarray}
    N_n&\equiv &S_{n+1}\nn\label{equivB}.
\end{eqnarray}
Such solutions exist classically within the Einstein-Cartan formalism or any formalism
allowing for degenerate metrics~\cite{Horowitz,Tom1}, but as we will comment later, our construction is more general.  

In such a setting, in the Euclidean region the wave function satisfies Schr\"odinger's equation (\ref{sch}), that is, (\ref{Sch1}) with:
\begin{eqnarray}
    X(b,n)&=& n\Delta  +  X (b,0)= n\Delta  -\frac{b^3}{3}+ kb
\end{eqnarray}
where $\Delta=\frac{4}{3}k^{3/2}$, $b\in (b_-,b_+)$ and $-\infty<n\le 0$ indexes the various spheres. This $X$ then connects with the Lorentzian $X=b^3/3+kb$ (valid for $b>0$) at $b=0$ and $n=0$ (see Fig.~\ref{figXn}). Crucially
\be
X(b_-,n)=X(b_+,n-1)
\label{solXT}
\ee
since the South pole of the $n$ sphere and North pole of the $n-1$ sphere have the same $X$.

The range of $X$ is now the whole real line, and with suitable fall-off conditions at infinity we have unitarity with inner product (\ref{innX}) just as for the Hartle-Hawking wave function (see Section~\ref{HH}). The wave functions are generic functions of the form $\psi(b,n,T)= F(X-T)$, such that:
\be
\int^\infty_{-\infty} dX\, |\psi|^2=\int^\infty_{-\infty}|\psi|^2\left|\frac{dX}{db}\right|db =1
\ee
but in the Euclidean region the wave functions have support only in $b\in (b_-,b_+)$. By construction the normalization is time independent. 
We can still ask what happens in the argument leading to the ``boundary'' terms in (\ref{Hherm})
as we jump from the North to the South of the next sphere, but this amounts to reading off from (\ref{solXT}) that:
\begin{equation}
    \psi(b_\pm,n,T)=\psi (b_\mp,n-1,T),
\end{equation}
so the boundary terms from contiguous spheres cancel out. Note that because of the determinant measure the probability density at any N or S is zero:
\begin{equation}
    {\cal P}(b_\pm,n,T)=|\psi|^2\left|\frac{dX}{db}\right|=0.
\end{equation}
This is very interesting, as it points to an unconventional transition between spheres as we now show.

Let us consider semiclassical states by choosing a Gaussian ${\cal A}$ centered on $\phi_0$ and with a spread $\sigma$. This leads to wave packets~\cite{sign,HH}:
\begin{widetext}
\be
\psi(b,n,T)=\frac{1}{(2\pi\sigma_T^2)^{1/4}}\exp\left[\frac{i}{\plk}\phi_0(X(b,n)-T)\right]\exp\left[-\frac{(X(b,n)-T)^2}{4\sigma_T^2}\right]
\ee
\end{widetext}
with $\sigma_T=\plk/(2\sigma)$.
The classical solution is depicted in Fig.\ref{figab} and combines (\ref{aE0}) and (\ref{bE0}) in the Lorentzian region with a repetition of (\ref{aE}) and (\ref{bE}) (restricted  so that $a>0$ is enforced\footnote{See the discussion in the next Section for what happens if we allow $a<0$.}) in the Euclidean region. 
For as long as the wave packets remain sharply (and single) peaked, they follow $\dot X=\dot T$ and, as we have seen, this is equivalent to the classical trajectory.

The semi-classical approximation, however,  badly breaks down near the transitions from one sphere to the next.
This is illustrated in Fig.~\ref{figprob}. What happens there is reminiscent of the singularity resolution in a radiation dominated Universe proposed in~\cite{GielenSing}. Here (as in~\cite{GielenSing}) the probability is always zero at the degenerate gluing point between spheres (at the classical singularity,  in~\cite{GielenSing}). As the classical trajectory approaches a North pole, we must therefore deviate from the semi-classical limit, since the peak of the probability cannot go there. What happens is that the probability peak stops its motion at a fixed distance to the North pole, dependent on $\sigma_T$.
As Fig.~\ref{figprob} shows, a symmetric peak then appears, equally distant from the South pole of the next sphere up the tower. As unimodular time progresses, the second peak grows at the expense of the first one. At the time when the classical trajectory would have reached the degenerate point, the 
wave function is actually a perfectly balanced superposition between these two peaks at non-degenerate points (middle panels of Fig.~\ref{figprob}). As time progresses the new peak comes to dominate and the old one is suppressed. As $T$ evolves by more than $\sigma_T$, the peak starts to move along the new sphere, from South to North pole, following the classical trajectory, until a new transition starts.

The fact that the N/S degenerate point is never reached seems to imply that this construction is more general than classical theories allowing for degenerate metrics. The transition between spheres may be seen as a truly quantum phenomenon.

Note that strictly speaking the action in this setting should be 
\begin{equation}
    S_0[b]=\Theta(b)S_L+ \Theta(- b,n=0) S_E[ b],
\end{equation}
instead of (\ref{Sgs}), since the counter $n$ appears in phase space. Another alternative is to use:
\begin{equation}
    S_0[b]=\Theta(T)S_L+ \Theta(- T) S_E[ b],
\end{equation}
but this falls within the remit of another set of theories currently being investigated. 

\section{Are there other options?}\label{other}

Within unimodular theory that does not seem to be the case. The obvious alternative, reflecting boundary conditions within a {\it single} sphere, presents a number of problems. One could use an ``images method'' for reflections off infinitely tall potential walls for implementing the boundary conditions in that case~\cite{interfreflex}. 
Recall that for a single infinite wall (crossing out $X<0$, say) one would take a normalized solution of the free unrestricted Schrodinger equation $\psi_U(X,T)$ and construct $\psi(X,T)=\psi_U(X,T)-\psi_U(-X,T)$ for $X>0$. This enforces the reflecting boundary condition, and $\psi$ is normalized in $X>0$, in spite of the interference terms. Adapting this to two infinite walls we could get solutions for the single Euclidean sphere (or half a sphere glued to a Lorentzian manifold). 

However, this would require accepting a theory with classical solutions $\dot T=\pm \dot X$, that is with waves traveling in both directions, in contradiction with the first order equation
(\ref{Sch1}). This is possible, but it would require extending the theory to accept $a<0$. 
The solutions (\ref{aE}) and (\ref{bE}) can then be extended to accept $\pm$ expansion factor (in analogy with the cosmological representation of a patch of AdS), and waves in $X$ and $T$ moving in two directions. A series of images can thus be built, describing the wave function being reflected back and forth between North and South pole (or just once off the South pole, if required). 

Two major flaws can be found. First, as already stated, the wave function would be reflected back to the Lorentzian domain, generating Hartle-Hawking packets, i.e. requiring both an expanding and a contracting Universe. One could consider several reflections, with the Lorentzian branches both in the $a>0$ of the theory, or with one in the $a<0$, but certainly we would not recover the Vilenkin asymptotic requirement. Second it is not clear that such a theory would be unitary. The inner product would have to be modified to accommodate solutions traveling in both directions. Perhaps with an entirely different construction this would be possible: namely a theory with a first order current.

\begin{figure}
\centering
\includegraphics[scale=0.135]{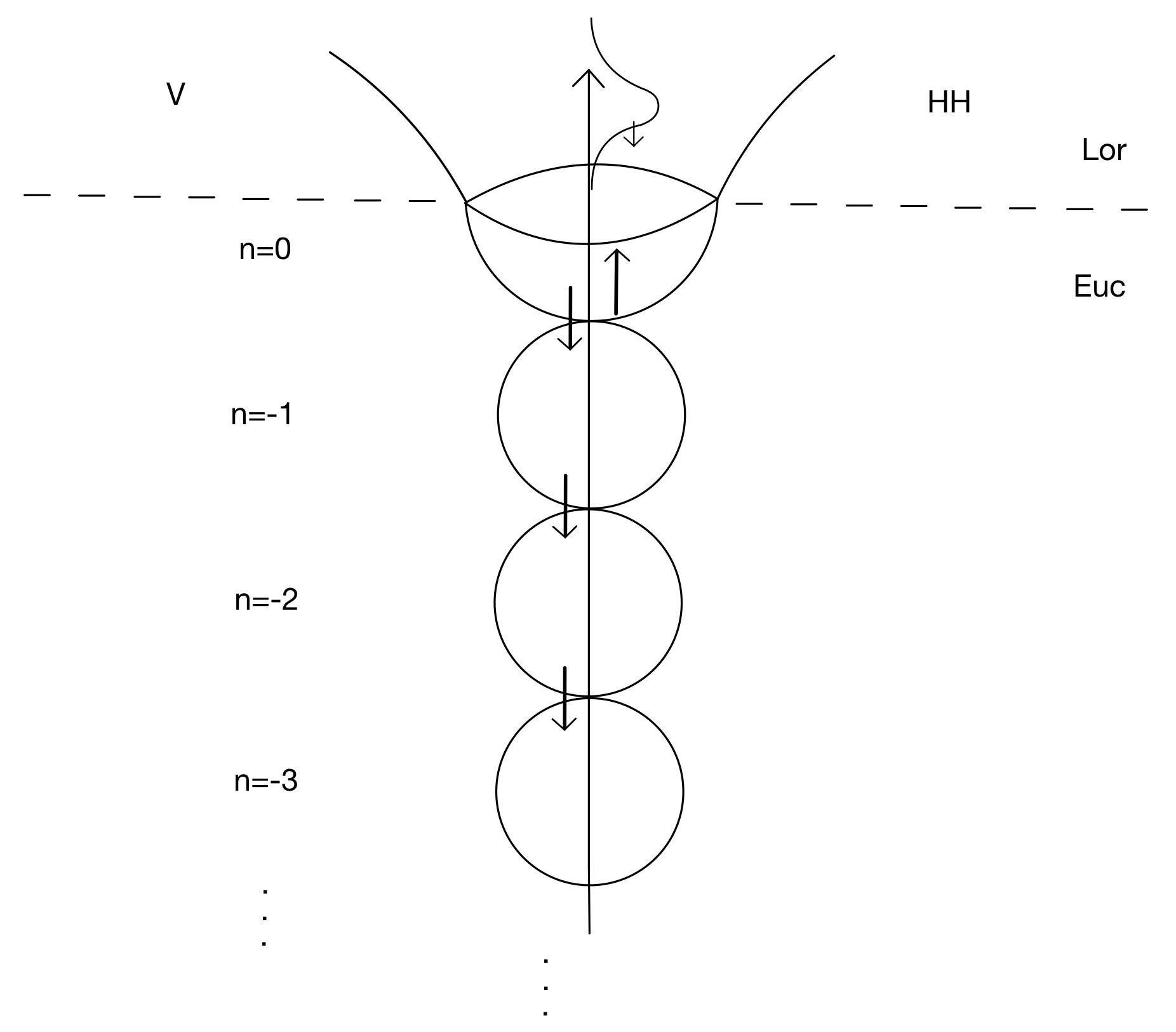}
\caption{The infinite tower of {\it real} Euclidean spheres supporting the Lorentzian Universe, required to maintain unitarity and present a single (outgoing) wave in the Lorentzian region. 
In this picture the arrows show motion as $T$ goes backwards. On the right we illustrate the effects of a reflection: it would only recreate the Hartle-Hawking aysmptotics (an incoming and an outgoing wave, at different unimodular times). On the left we illustrate the infinite Sisyphus sequence of cycles, required for Vilenkin asymptotics.} 
\label{figuni}
\end{figure}

Given that we would gain nothing from this complication (i.e. we would be back to Hartle-Hawking Lorentzian conditions), we will not explore this possibility further in this paper. But note that the the fact that we have a reflection within the same sphere is physically different from having multiple distinct spheres. The former leads to interference between reflected and incident waves (revealing the internal beats of packets in the probability) near the reflecting pole and a Hartle-Hawking packet in the Lorentzian region; the latter has no interference and we can indeed have only outgoing packets.

\section{Conclusions}

In this paper we have tried and failed to implement unitarity for the Vilenkin wave function in the unimodular extension of gravity. The failure resides squarely in the connection's detour into the imaginary domain, associated with the Vilenkin boundary condition. This is true for the monochromatic partial waves (with fixed $\Lambda$) and contaminates the construction of wave packets in unimodular theory. The usual setup for a conserved inner product (so successful for packets of Hartle-Hawking wave functions) then fails. Attempts to use the Laplace transform in lieu of the Fourier integral only force the probability to live on the Hartle-Hawking connection contour. More structurally, the problem is that imaginary connections lead to anti-Hermitian Hamiltonians and so anti-unitarity. This is the unavoidable consequence of seeing the Euclidean theory (wherein the ``nothing'' lies) as a complex image within a Lorentzian theory. Time, connection and Chern-Simons functional then necessarily become imaginary, and unitarity is lost. 

Without prejudice to the work of~\cite{HHpackets} (where none of these complications are found for the Hartle-Hawking state) or the stated intentions of~\cite{Vil-interpretation} (to regard unitarity as a mere approximation subject to breaking down), in this paper 
we put forward an alternative. The connection detour into the imaginary domain, usually seen as an Euclidean image within a Lorentzian theory, could also be interpreted as an actual transition from Euclidean to Lorentzian signature within a fundamentally real theory. If that is the case, all variables, action and Hamiltonian remain real, so that no structural obstacles to unitarity are found.

Nonetheless, the question still arises as to what happens at the boundary (the South pole) of the (now real) half-sphere. We proposed that a way to enforce unitarity whilst keeping a single (outgoing) wave in the Lorentzian domain is to allow the wave function to travel through a tower of spheres for an eternity in unimodular time. Such solutions do exist in gravity formulations allowing for degenerate metrics, but we stress that we could envisage our proposal as a purely quantum construction, required by unitarity and Vilenkin asymptotics. We found that semiclassical solutions with a peak following the classical equations become essentially quantum at the degenerate gluing points. Hence our proposal could even apply to standard GR, but only quantum mechanically. 

We may call this proposal ``Sisyphus boundary conditions'', since it is a time-reversal of Sisyphus' ordeal. Sysiphus was condemned to roll a ball uphill only for it roll down restarting the cycle for the whole of a future eternity. In our proposal the wave function climbs up and down an infinite tower of distinct spheres for a past eternity. In the same way that Sisyphus' punishment avoids death, ours avoids birth, i.e. a moment of cosmic creation. Ultimately this is the reason why unitarity is kept, whilst complying with Vilenkin's boundary conditions in the Lorentzian region. 

We close with a few comments on the relation between our proposal and work found in the literature. One objection to a tower of spheres is that should one try to put a general quantum field theory on this background, then the action cannot be made to converge, due to the gluing points~\cite{Witten,JeanLuc} (more generally one has to face the instabilities unveiled in~\cite{JL1,JL2,JL3}). 
But as we saw, the degenerate points are never reached by the peak of the wave function, with the evolution at transition unimodular times being intrinsically quantum. So the concept of quantum fields on top of a classical background breaks down at these points. In addition the evaluation of the path integral for the background assumes that the metric (and presumably the connection) is complexified to bypass such points, something that does not happen here. Another objection targets the whole concept of semi-classical time in the first place, as suggested in~\cite{notime}. Are these two facts sufficient to allay the concerns of~\cite{JL1,JL2,JL3,Witten,JeanLuc}? A translation between canonical and the path integral approach is non-trivial, and what is meant by "contour", "measure", "convergence", etc is strictly speaking not the same. For this reason it is hard to see what the implications are, for example, for the concept of ``fuzzy instantons'' (e.g.~\cite{fuzzy}). A full evaluation of this issue from the path integral point of view is left to future work. 

We finally stress that our proposal is fundamentally different from postulating a cyclic Chern-Simons functional, $X$, and so a cyclic unimodular time, $T$. We could consider a cyclic unimodular time, $T_c$, related to our $T$ by:
\be
T=n\Delta  + (-1)^n T_c
\ee
with $\Delta=\frac{4}{3}k^{3/2}$ and $-\Delta/2 <T_c<\Delta/2$. The fact that our $T$ is built from a cyclic process plus a integer counter, $n$, should not be seen as an anomaly. Every practical clock is the result of a periodic process (the oscillation of a pendulum, the vibration of a crystal, etc) plus a counter. Our timing  system (in years, weeks, etc) also ``wraps" around, triggering a counter. Since our time is not imaginary, there is no good reason to make it cyclic.

\section{Acknowledgments}
We thank Steffen Gielen, Jean-Luc Lehners and Alex Vilenkin for discussions related to this paper. This work was supported by FCT Grant No. 2021.05694.BD (B.A.) and the STFC Consolidated Grant ST/T000791/1 (J.M.).


\end{document}